\documentclass[lettersize,journal]{IEEEtran}
\usepackage{amsmath,amsfonts}
\usepackage{algorithmic}
\usepackage{algorithm}
\usepackage{array}
\usepackage[caption=false,font=normalsize,labelfont=sf,textfont=sf]{subfig}
\usepackage{textcomp}
\usepackage{stfloats}
\usepackage{url}
\usepackage{verbatim}
\usepackage{graphicx}
\usepackage{cite}
\usepackage{booktabs}
\usepackage{tcolorbox}
\usepackage{adjustbox}

\usepackage[utf8]{inputenc}
\usepackage[T1]{fontenc}
\usepackage{tikz}
\usetikzlibrary{
    shapes.geometric,
    arrows.meta,
    positioning,
    shadows
}

\usepackage{xspace}
\newcommand{\ie}{i.e.\@\xspace} \newcommand{\eg}{e.g.\@\xspace}  \newcommand{\etal}{\textit{~et~al.\@}\xspace}
\usepackage{url}

\usepackage{pgfplots}
\usepackage{tabularx}
\pgfplotsset{}\usepgfplotslibrary{statistics}
\usepgfplotslibrary{groupplots}

\begin{document}

\title{Obfuscating Code Vulnerabilities against Static Analysis in JavaScript Code}

\author{Francesco Pagano,~\IEEEmembership{Member,~IEEE,} 
        Lorenzo Pisu,~\IEEEmembership{Member,~IEEE,}
        Leonardo Regano,~\IEEEmembership{Member,~IEEE,}
        Davide Maiorca,~\IEEEmembership{Member,~IEEE,} 
        Alessio Merlo,~\IEEEmembership{Senior Member,~IEEE,} 
        Giorgio Giacinto,~\IEEEmembership{Senior Member,~IEEE} 
        
\thanks{Manuscript received December 1, 2025. Francesco Pagano and Lorenzo Pisu share dual first authorship. \textit{(Corresponding author: Leonardo Regano.)}}\thanks{Francesco Pagano is with the Department of Computer Science, University of Verona, Verona, Italy (e-mail: francesco.pagano@univr.it).}
\thanks{Lorenzo Pisu, Leonardo Regano, and Davide Maiorca are with the Department of Electrical and Electronic Engineering, University of Cagliari, Cagliari, Italy (e-mail: lorenzo.pisu@unica.it; leonardo.regano@unica.it; davide.maiorca@unica.it).}
\thanks{Alessio Merlo is with the School of Advanced Defense Studies (CASD), Rome, Italy (e-mail: alessio.merlo@unicasd.it).}
\thanks{Giorgio Giacinto is with the Department of Electrical and Electronic Engineering, University of Cagliari, Cagliari, Italy, and with the National Interuniversity Consortium for Informatics, Rome, Italy (e-mail: giorgio.giacinto@unica.it).}
\thanks{This work was partially supported by project SERICS (PE00000014) under the NRRP MUR program funded by the EU - NGEU.}
}

\markboth{IEEE TRANSACTIONS ON INFORMATION FORENSICS AND SECURITY}{Pagano \MakeLowercase{\textit{et al.}}: Obfuscating Code Vulnerabilities against Static Analysis in JavaScript Code}

\maketitle

\begin{abstract}
Code obfuscation is widely adopted in modern software development to protect intellectual property and hinder reverse engineering, but it also provides attackers with a powerful means to conceal malicious logic inside otherwise legitimate JavaScript code. In a software supply chain where a single compromised package can affect thousands of applications, this raises a critical question: how robust are the Static Application Security Testing (SAST) tools that CI/CD pipelines rely on as automated security gatekeepers?
This paper answers that question by empirically quantifying the impact of JavaScript obfuscation on state-of-practice SAST. We define a realistic supply-chain threat model in which an adversary injects vulnerable code and iteratively obfuscates it until the pipeline reports a clean scan. To measure the resulting degradation, we introduce the Vulnerability Detection Loss (VDL) metric and conduct a two-phase study. First, we analyze 16 vulnerable-by-design Node.js web applications from the OWASP directory; second, we extend the analysis to 260 in-the-wild JavaScript/Node.js projects from GitHub. Across both datasets, we apply eight semantics-preserving obfuscation techniques and their combinations and evaluate two representative SAST tools, Njsscan and Bearer. Even a single obfuscation technique typically suppresses most baseline findings, including high-severity issues, while stacking techniques yield near-total evasion, with VDL often approaching 100\%. Our results show that current JavaScript SAST is fundamentally not robust against commonplace obfuscations and that “clean” reports on obfuscated code may offer only a false sense of security. Finally, we discuss practical mitigation guidelines and directions for obfuscation-aware analysis.

\end{abstract}

\begin{IEEEkeywords}
Code Obfuscation \and Static Analysis \and SAST \and Web Security \and Javascript Vulnerabilities
\end{IEEEkeywords} \section{Introduction}
JavaScript is a fundamental component of the current web due to its widespread use in client-side and server-side development. This central role makes the JavaScript ecosystem, particularly its package managers like npm, a prime target for software supply chain attacks, where a single compromised library can impact thousands of downstream projects. Most of the most widely used web development frameworks are based on JavaScript, according to~\cite{webFrameworkPercentageUsage}. Code obfuscation is a defensive method frequently used by developers to safeguard their intellectual property, but it can also be used as a weapon by malicious actors. Obfuscation is a cloaking technique that allows attackers to hide security flaws by changing code into a syntactically complex but functionally equivalent form. Attackers can hide a wide range of critical flaws, such as those enabling cross-site scripting (XSS), data exfiltration, or remote code execution, directly within the logic of otherwise functional code. The seriousness of this threat has been proven in other fields. For instance, a recent study by Pagano et al.~\cite{pagano2024obfuscating} empirically demonstrated that obfuscation significantly reduces the ability of SAST tools to detect vulnerabilities in Android applications. Driven by these results, this paper offers the first extensive empirical analysis to determine whether this threat is as severe in the JavaScript ecosystem. JavaScript presents particular challenges for static analysis because of its dynamic nature and variety of execution environments, which makes this a crucial field of study. These challenges stem from features like weak typing, prototype-based inheritance, and highly asynchronous, event-driven control flows, which already complicate the ability of static analyzers to build accurate models of program behavior even before obfuscation is applied.

Discovering vulnerabilities in complex codebases requires the use of automated security testing tools. Static Application Security Testing (SAST) and Dynamic Application Security Testing (DAST) are the two main categories into which these tools fall. Both approaches are often used in tandem as part of a comprehensive DevSecOps strategy to provide layered security. In order to find known vulnerability signatures, SAST tools examine an application's source code directly for insecure code patterns and data flows. This 'shift-left' approach allows for the early detection of issues, reducing the cost and complexity of remediation before the code is merged or deployed. DAST tools, on the other hand, test an application that is currently running from the outside, keeping an eye on its inputs and outputs to find security vulnerabilities in its behavior, regardless of the structure of the underlying source code. Furthermore, DAST's effectiveness is contingent on the quality and coverage of the test cases used to exercise the application's features, potentially leaving untested code paths un-analyzed. Because SAST tools' analysis is essentially reliant on the source code's syntactic structure, this paper focuses only on them. As a result, they are actually the most vulnerable to obfuscation techniques, which modify the appearance of code without affecting its behavior during execution.

Our study is the first to investigate this issue by empirically analyzing the impact of code obfuscation techniques on SAST tools. In particular, our study aims to answer the following research questions:
\begin{enumerate}
\item[\textbf{RQ1:}] What is the baseline performance of SAST tools on vulnerable-by-design websites, and how does this performance degrade by obfuscating their JavaScript code with one or multiple obfuscation techniques?
\item[\textbf{RQ2:}] What is the individual impact of different obfuscation techniques on the vulnerability detection capabilities of SAST tools, when analysing real-world websites?
\item[\textbf{RQ3:}] What is the cumulative effect of stacking multiple obfuscation techniques on the vulnerability detection capabilities of SAST tools, when analysing real-world websites?
\end{enumerate}

To answer such research questions, we conducted a thorough, two-phase empirical assessment designed to measure the impact of obfuscation on SAST tools. This comprehensive methodology was chosen to ensure that our conclusions are robust and generalizable. We first examined a set of deliberately vulnerable benchmark applications, provided by the OWASP Foundation~\cite{owaspVulnerableWebApps}, to create a trustworthy ground truth. To this dataset, we methodically applied every possible combination of obfuscation techniques from a chosen, representative obfuscation tool. The use of a ground-truth dataset with confirmed vulnerabilities is critical, as it enables the precise calculation of the false negative rate introduced by each obfuscation strategy, a key metric for evaluating detection loss. This made it possible for us to accurately calculate the detection loss for every SAST tool in a controlled setting with respect to each distinct obfuscation technique. Second, we used the same set of obfuscation combinations on an extensive set of 260 applications that we obtained from GitHub to validate these results at scale. Analyzing these 'in-the-wild' applications is essential for ecological validity, as they contain the architectural complexities, idiomatic code, and third-party dependencies that are often absent from academic benchmarks, but are commonplace in production systems. By showing how our findings affect the actual code, this dual approach guaranties that our conclusions are both externally relevant and internally valid, as determined by the ground-truth analysis. Finally, our results benchmark the resilience of existing-generation SAST tools and give a quantitative measure of the risk obfuscation poses to automated security pipelines. For reproduction purposes, we open-source in a GitHub repository\footnote{\url{https://github.com/X3no21/JavascriptObfuscationSAST}} the datasets and scripts we used to perform our empirical assessment.

The remainder of this paper is organized as follows. Section~\ref{sec:related} reviews related work on the empirical evaluation of SAST tools and on code obfuscation as both a defensive and offensive technique. Section~\ref{sec:background} introduces the technical background on JavaScript SAST and the obfuscation strategies considered in our study. Section~\ref{sec:model}  formalizes the threat model and describes the attacker’s workflow for poisoning software supply chains through obfuscated vulnerabilities. Section~\ref{sec:assessment} details the assessment methodology, including tool selection, metrics, and experimental factors, while Section~\ref{sec:experiments} presents the design and results of our experimental campaign on both ground-truth and in-the-wild datasets. Section~\ref{sec:discussion} discusses the implications and limitations of our findings, and Section~\ref{sec:conclusions} concludes the paper and outlines directions for future research. \section{Related works}
\label{sec:related}

Static Application Security Testing (SAST) for web code has been extensively studied from both methodological and empirical angles. Empirical evaluations consistently show that, even on real-world benchmarks,  mainstream SAST tools struggle to surface a large fraction of true vulnerabilities. For instance, Li et al. \cite{li2023sastjava} systematically compare state-of-the-art SAST tools for Java code (including CodeQL, Semgrep, and SonarQube) and reported limited recall even when multiple tools are combined. Focusing on JavaScript and Node.js, Brito et al. \cite{brito2023sastjs} built a curated dataset of confirmed npm vulnerabilities and found that nine popular analysers (including NodeJsScan, ODGen, and Microsoft's DevSkim) miss many vulnerabilities, including those included in the OWASP Top 10, and show a high false positive rate. 
On the methodological side, Ferreira et al. \cite{ferreira2024sastjs} recently introduced Graph.js\footnote{We did not consider this tool for our evaluation since it is currently unable to analyze websites, but only npm packages.}, a SAST tool for JavaScript that leverages Multiversion Dependency Graphs to detect four types of vulnerabilities on server-side npm dependencies: prototype pollution, OS command injection, arbitrary code execution, and path traversal. However, they do not provide any assessment of the adversarial robustness of their approach. Collectively, these works evaluate either the effectiveness of existing analysis methods on real code or propose stronger ones. None of them, however, quantifies how obfuscation alters SAST analysis outcomes for JavaScript web code, which is the focus of our study.

Obfuscation is widespread on the web, as reported by large-scale measurement studies. Sarker et al. \cite{sarker2020jsobf} assessed the concealed usage of browsers' APIs at scale, reporting that 95\% of the Alexa top 100K domains use such concealed API calls. Moog et al. \cite{moog2021statically} statically detected that 90\% the Alexa top 10K domains employed some sort of JavaScript obfuscation techniques, evidencing that JavaScript is routinely obfuscated in real-world settings. These measurements justify studying robustness to obfuscation for any static defence integrated in CI/CD, including SAST.

Closest to our adversarial angle are works that examine the impact of obfuscation on other automated analyses. In the Android domain, Pagano et al.~\cite{pagano2024obfuscating} demonstrated that deliberately obfuscating vulnerable code can significantly reduce the vulnerability detection rates of SAST tools for mobile applications. Furthermore, several studies in malware detection show that obfuscation degrades the reliability of static features and machine learning pipelines: for example, Molina-Coronado et al. \cite{molinacoronado2025androidmalwareobf} analysed how standard Android obfuscations perturb static features used by detectors. To tackle this issue, obfuscation-resilient detection approaches for Android malware have been proposed, such as MosDroid by Sharma et al. \cite{sharma2025androidmalware}, which leverages multiple machine learning models trained on multisets of malware opcode sequences. Concerning web code, Ren et al. \cite{ren2023obfJSmalware} showed that common JavaScript obfuscators systematically degrade the performances of static ML-based malicious code detectors.

A complementary line of research investigates countermeasures to transformed JavaScript. A recent example is JSimpo~\cite{chen2025jsimpo}, a structural deobfuscation framework for JavaScript code. The authors indicate that preliminary deobfuscation may be beneficial before analyzing malicious JavaScript code. Indeed, the same approach may be adopted prior to analyzing obfuscated third-party code with SAST tools.

Summarizing, while the SAST effectiveness on real software have been extensively researched, and despite the widespread employment of obfuscation for JavaScript code,
to the best of our knowledge, no prior work systematically quantifies how JavaScript code obfuscation impacts the detection capabilities of mainstream SAST tools used in DevSecOps for web projects. Our study fills this gap with a controlled, large-scale assessment, thereby complementing current literature on the impact of obfuscation on static code analysis tools. \section{Background}
\label{sec:background}

This section provides the necessary background on JavaScript SAST tools, common web application vulnerabilities, and code obfuscation techniques employed in our study.

\subsection{SAST Techniques}

In SAST, software is analyzed by looking at either the compiled code (black-box mode) or the source code (white-box mode). In the latter instance, the software's logic is first represented through the use of reverse engineering. JavaScript lacks a separate compilation step that creates binary executables, though, because it is an interpreted language. For this reason, SAST for JavaScript only functions in white-box mode, which depends on the source code being available. This requirement is always satisfied for client-side JavaScript because the code is accessible straight from the developer tools of any web browser. However, the availability of server-side JavaScript is based upon the project being open-source or an analyst having authorized access to the web server's repository. JavaScript SAST tools use a number of complex techniques, frequently in combination, to identify vulnerabilities.

\paragraph{Analysis of Abstract Syntax Tree (AST)} The majority of analyses begin with the parsing of source code into an Abstract Syntax Tree (AST). A parser performs this process, converting the code's raw text into a hierarchical tree structure in which each node stands for a construct found in the source code, such as an if-statement, function declaration, or variable assignment. Because it enables tools to comprehend the grammar of the code and the relationships between elements, going beyond basic text-based searches, the AST is the main data structure upon which more complex analyses are constructed.

\paragraph{Pattern-Based Detection} Pattern-based detection is the most straightforward use of AST analysis. Using the AST, this method looks for certain structural patterns, unsafe function calls (like using `eval()` or `innerHTML`), or out-of-date library usages that are frequently linked to known vulnerabilities. It functions similarly to an advanced linter, identifying potentially unsafe constructs with a predetermined set of rules and signatures. This approach is quick and efficient for finding "low-hanging fruit," but it lacks contextual awareness and is unable to determine whether a flagged pattern is truly exploitable, which can result in a high number of false positives.

\paragraph{Control Flow Analysis} SAST tools carry out Control Flow Analysis to comprehend the behavior of the program beyond static patterns. This entails using the AST to create a Control Flow Graph (CFG), in which directed edges stand in for control flow transfers (such as conditionals, loops, or function calls) and nodes for simple blocks of straight-line code. The tools can detect insecure program flows, unreachable dead code, and logical errors by examining this graph. It can be used, for instance, to confirm that a required authorization or authentication check on all potential execution paths leading to a sensitive operation always comes before the operation itself.

\paragraph{Taint Analysis} Taint analysis focuses on the data that pass through those paths, whereas control flow analysis charts the logic of the program's execution. This effective method works in three steps: first, it flags all information from unreliable external \textit{sources} (like user input from a URL parameter) as "tainted." Second, it spreads this taint when the data is used in operations or allocated to other variables. Third, it determines whether any contaminated data makes it to a sensitive \textit{sink} (such as a database query or a command execution function) without first going via a sanitization function. The tool highlights a potential vulnerability, such as SQL Injection or Cross-Site Scripting (XSS), if it detects a corrupted data flow.

\paragraph{Symbolic Execution} Some advanced tools use symbolic execution for more comprehensive and in-depth analysis. This approach uses symbolic variables to explore program paths rather than concrete data values, methodically producing mathematical constraints that must be met to follow each path. It is very effective in finding complicated bugs, such as path traversal or integer overflow vulnerabilities, because it can use a constraint solver to identify any specific input values that might cause the program to enter a vulnerable state. Nevertheless, this method meets the problem of "path explosion" in complicated applications, and, in order to remain feasible, it is frequently coupled with concrete execution (a hybrid strategy called concolic execution).

\subsection{Web Code Obfuscation}
\label{sec:background:obfuscation}
Code obfuscation is an umbrella term encompassing various code transformation techniques intended to significantly increase the effort needed by an attacker to reverse engineer a target application. Collberg \etal~\cite{collberg1997taxonomy,collberg2002watermarking} subdivide such techniques in four broad categories: data obfuscation (\eg encoding or hiding literals and variable runtime values), layout obfuscation (\eg changes in the source code formatting, renaming of objects identifiers), control obfuscation (\eg alterations to the source code to hinder attacker comprehension of the program's runtime control flow), and preventive obfuscation (runtime checks and reactions to detect and disrupt attacker tools, \eg decompilers or debuggers). Such transformations are typically applied to code by means of automatic tools (\ie code obfuscators). Outside the web domain, both commercial and open-source tools are available to obfuscate source code written in various programming languages, for example C/C++ (Tigress, Obfuscator-LLVM, Stunnix C/C++ Obfuscator) and Java (dProtect, JObfuscator, Allatori Java Obfuscator). Similarly, in the web domain, various open-source and commercial obfuscators are available for Javascript code. Most open-source obfuscators, such as Terser, UglifyJS, Closure Compiler and SWC, operate primarly as minifiers or optimizers, performing transformations belonging to the data and layout obfuscation categories, for example changing variable names to unintelligible ones and removing whitespaces and comments. A notable exception is javascript-obfuscator, which, to the best of our knowledge, is the only available open-source tool supporting techniques from all the four categories from Collberg's taxonomy. Similar commercial tools exist (\eg PreEmptive JSDefender, Digital.ai Application Security), but we exclude those from our experimentation due to licensing costs. 
Thus, we selected javascript-obfuscator as the subject of our experimentation. Following, we briefly describe the javascript-obfuscator techniques that we employ in our study:

\begin{itemize}
    \item \emph{Compact (CMP)}: a layout obfuscation that condenses all code in a single line and removes all whitespaces, thus hindering manual inspection of the obfuscated code;
    \item \emph{Control Flow Flattening (CFF)}: first introduced by Wang \etal~\cite{wangFlatteningTechReport}, this control obfuscation hides the application's control flow by placing all the code in a loop/switch statement; each basic block of the original code is placed in a case of the switch statement; the original application flow is preserved at runtime by properly updating the switch variable at runtime, conversely making harder its reconstruction only with static analysis means;
    \item \emph{Dead Code Injection (DCI)}: a control obfuscation that introduces blocks of dead code to increase the application's \textit{critical section}, \ie the source code lines that the attacker must analyse to successfully reverse engineer the targeted application; to harden the identification of dead code blocks, those are inserted in conditional statements using Opaque Predicates~\cite{collbergOpaque}, \ie obfuscated tautological predicates;
    \item \emph{Debug Protection (DP)}: a preventive obfuscation~\cite{muschAntidebugging} used to disrupt the use of debuggers, \eg the ones included in the Developer Tools of modern browsers, by repeatedly triggering breakpoints in various points of the protected code with the \texttt{debugger} JavaScript statement;
    \item \emph{Self Defending (SD)}: a preventive obfuscation that hinders the use of code beautifiers to restore the original form of code previously compacted through the CMP obfuscation; when the application starts, this technique will perform a runtime check that will input the protected code into a regex function that, in presence of a whitespace (which are removed by the CMP obfuscation), will cause catastrophic backtracking, causing the beautified code to hang indefinitly;
    \item \emph{Simplify (SIMP)}: a layout obfuscation that inlines multiple lines of code where allowed by the JavaScript syntax to harden manual code inspection, \eg declaration and assignment of multiple variables, inlined conditional statements;
    \item \emph{Split Strings (SS)}: a data obfuscation that hides string literals by splitting them into chunks of fixed length, in order to harden the search for a specific string in the obfuscated code; to preserve the application's business logic, the split strings are reconstructed at runtime through concatenation;
    \item \emph{String Array (SA)}: a data obfuscation that, as the previous obfuscation, hides string literals in the obfuscated code, in this case by placing them in a dedicated array, and replacing the removed literals with lookups of the aforementioned array (index-based or via a dedicated decoder function), so each string is retrieved at runtime instead of appearing in cleartext.
\end{itemize}

Table~\ref{tab:obfuscation_techniques} summarizes the selected obfuscation strategies and their corresponding types.

\begin{table}[!ht]
    \centering
    \small
    \begin{tabular*}{\columnwidth}{@{\extracolsep{\fill}}lll}
        \toprule
        \textbf{Acr.} & \textbf{Technique Name} & \textbf{Type} \\
        \midrule
        CMP & Compact & Layout \\
        CFF & Control Flow Flattening & Control \\
        DCI & Dead Code Injection & Control \\
        DP & Debug Protection & Preventive \\
        SIMP & Simplify & Layout \\
        SA & String Array & Data \\
        SD & Self Defending & Preventive \\
        SS & Split Strings & Data \\
        \bottomrule
    \end{tabular*}
    
    \vspace{5pt}
    \caption{Summary of the selected obfuscation techniques employed in the study.}
    \label{tab:obfuscation_techniques}
\end{table}

 \section{Threat Model}
\label{sec:model}

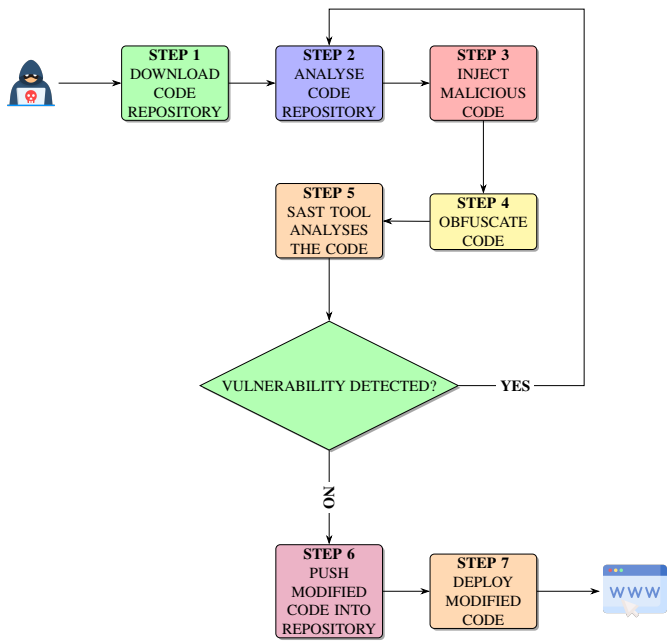
\begin{figure}
    \centering
    \begin{adjustbox}{width=\columnwidth,keepaspectratio}
        \begin{tikzpicture}[font=\Large]
            \tikzset{
              block/.style = {rectangle, rounded corners, draw=black, text width=9em, text centered, minimum height=3.5em,
                              drop shadow={opacity=0.7, shadow xshift=2pt, shadow yshift=-2pt}, font=\Large},
              decision/.style = {diamond, draw=black, fill=green!30, text centered, minimum size=2.5em, aspect=2,
                                 drop shadow={opacity=0.7, shadow xshift=2pt, shadow yshift=-2pt}, font=\Large},
              actor/.style = {text centered, font=\Large},
              arrowstyle/.style = {draw, -{Stealth[length=3mm, width=2mm]}, thick},
              edge_label/.style = {font=\Large\bfseries, sloped, midway}
            }
            
            \node[actor] (attacker) {\includegraphics[width=1.5cm]{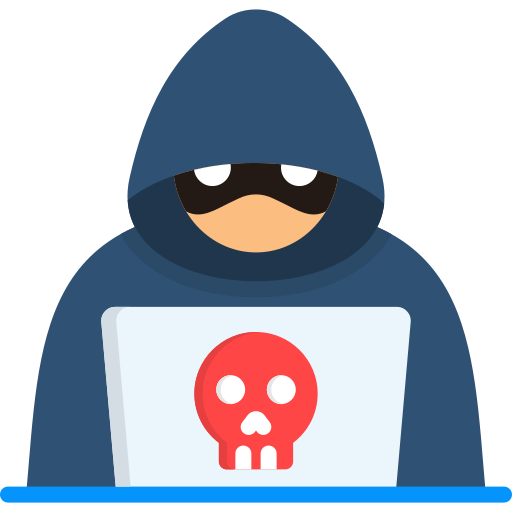}};
            \node[block, fill=green!30, right=2cm of attacker] (step1) {\textbf{STEP 1}\\ DOWNLOAD CODE REPOSITORY};
            \node[block, fill=blue!30, right=1.5cm of step1] (step2) {\textbf{STEP 2}\\ ANALYSE CODE REPOSITORY};
            \node[block, fill=red!30, right=1.5cm of step2] (step3) {\textbf{STEP 3}\\ INJECT MALICIOUS CODE};
            \node[block, fill=yellow!40, below=2.33cm of step3] (step4) {\textbf{STEP 4}\\ OBFUSCATE CODE};
            \node[block, fill=orange!30, below=2cm of step2] (step5) {\textbf{STEP 5}\\ SAST TOOL ANALYSES THE CODE};
            \node[decision, below=2cm of step5] (decision) {VULNERABILITY DETECTED?};
            \node[block, fill=purple!30, below=3.0cm of decision] (step6) {\textbf{STEP 6}\\ PUSH MODIFIED CODE INTO REPOSITORY};
            \node[block, fill=orange!30, right=1.5cm of step6] (step7) {\textbf{STEP 7}\\ DEPLOY MODIFIED CODE};
            \node[actor, right=2cm of step7] (www) {\includegraphics[width=2cm]{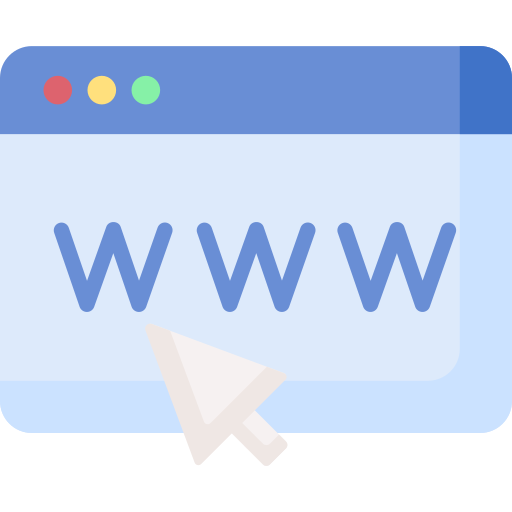}};
            
            \path [arrowstyle] (attacker) edge (step1);
            \path [arrowstyle] (step1) edge (step2);
            \path [arrowstyle] (step2) edge (step3);
            \path [arrowstyle] (step3) edge (step4);
            \path [arrowstyle] (step4) edge (step5);
            \path [arrowstyle] (step5) edge (decision.north);
            \path [arrowstyle] (step6) edge (step7);
            \path [arrowstyle] (step7) edge (www);
            \path [arrowstyle] (decision) edge node[edge_label, right, xshift=-5mm, fill=white] {NO} (step6);
            \draw[arrowstyle] (decision.east) -- node[edge_label, pos=0.3, right, fill=white] {YES} ++(4,0) |- ++(0,11cm) -- ++(0,1cm) -| (step2.north);
        \end{tikzpicture}
    \end{adjustbox}
    \caption{Threat Model}
    \label{fig:threat_model}
\end{figure}

This section outlines a threat model demonstrating how code obfuscation can be exploited to circumvent SAST in current software development campaigns. The widespread adoption of languages such as JavaScript throughout the development stack, where a substantial majority of contemporary web frameworks depend on it~\cite{webFrameworkPercentageUsage}, has highlighted the essential necessity of effective SAST. Nonetheless, these tools can be evaded via intentional software supply chain attacks. Such attacks can appear differently based on the repository's characteristics. In private repositories, an attacker may exploit compromised credentials or function as a malicious insider. In public, open-source repositories, an adversary may masquerade as a legitimate contributor, gaining the community's trust through authentic contributions prior to injecting malicious code. In both scenarios, obfuscation serves as an effective method for hiding these malicious modifications, as outlined in the threat model illustrated in Figure~\ref{fig:threat_model}.

The attack sequence starts with the initial breach phase, resulting in the attacker downloading the source code (Step 1). In a private repository context, access is usually obtained through illicit methods, including targeted phishing campaigns directed at developers, exploiting misconfigurations within the code hosting platform, or using stolen credentials. Alternatively, it might come from an insider threat. In public, open-source repositories, the strategy is more complex and depends on social engineering: the attacker develops a reputation as a trustworthy contributor by gradually submitting legitimate bug fixes or minor features, ultimately leveraging the established trust to propose a malicious modification. After gaining access to the codebase, the adversary conducts a comprehensive analysis to understand its architecture and identify a suitable location for the injection of a malicious payload (Step 2). Upon identifying an appropriate location, the malicious code is integrated into the original source (Step 3). The key component of the evasion strategy lies in the subsequent obfuscation of the modified code (Step 4), which converts it into a functionally equivalent yet syntactically complicated version intended to confuse SAST engines and human reviewers. The validation loop (Step 5) is a crucial stage in which the attacker evaluates their obfuscated code using the same SAST tool employed by the project maintainers, progressively enhancing the obfuscation until the vulnerability is undetectable. The attacker uploads the poisoned malicious code to the repository after successfully evading the SAST tool (Step 6). This can happen by submitting an apparently safe pull request in an open-source project or by making a direct commit in a compromised private repository. Lastly, this action usually starts a Continuous Integration/Continuous Deployment (CI/CD) pipeline in modern workflows, which automatically deploys the compromised code to the production environment (Step 7). The development workflow has now been effectively turned into an attack vector, allowing the attacker to activate the hidden exploit once the application has been compromised. \section{Assessment Methodology}
\label{sec:assessment}
In this section, we describe our assessment methodology, providing details on how we selected the SAST tools for our study.

\subsection{Selection Criteria for SAST Tools}

The selection of appropriate SAST tools for this study was guided by a set of well-defined criteria to ensure our evaluation is relevant, reproducible, and focused. A rigorous selection framework is essential for generating findings that are applicable to the security practices used in real-world development environments. The tools included in our analysis were therefore required to meet the following prerequisites:

\begin{itemize}
    \item \textbf{JavaScript Support:} The tool must provide robust support for modern JavaScript (ECMAScript 6 and later), with the ability to analyze both client-side browser code and server-side Node.js applications.

    \item \textbf{Maturity and Adoption:} The tool must be well-established and widely used within either the open-source community or commercial development pipelines. This ensures that our findings are applicable to tools that developers and organizations currently rely on for security assurance.

    \item \textbf{Broad Vulnerability Coverage:} The tool must be capable of detecting a wide range of common security vulnerabilities, such as those outlined in the OWASP Top 10 for web applications, including Cross-Site Scripting (XSS), Injection flaws, and Insecure Deserialization.

    \item \textbf{Accessibility:} The tool must be available for academic research, either through an open-source license, a free tier, or a trial license that is sufficient for conducting the full scope of our experiments.
\end{itemize}

\subsection{Analyzed SAST Tools}

Based on the criteria outlined above, we selected a portfolio of five tools. This selection represents a cross-section of approaches to static analysis in the JavaScript ecosystem, from comprehensive platforms to specialized open-source tools.

\paragraph{SonarQube~\cite{repoSonarQube}} As one of the most widely adopted platforms for continuous code quality and security, SonarQube serves as an industry benchmark. It integrates seamlessly into CI/CD pipelines to function as a "quality gate," preventing code that does not meet security and quality standards from being deployed. Its analysis engine for JavaScript is multi-faceted, combining traditional lexical and pattern-based checks with a more sophisticated data-flow analysis engine capable of detecting complex injection vulnerabilities. Its security rules are often aligned with established standards like the OWASP Top 10 and Common Weakness Enumeration (CWE), making its inclusion representative of a mature, all-in-one security platform.

\paragraph{Bearer~\cite{repoBearer}} Bearer is an open-source SAST tool with a strong focus on data security and privacy risks, in addition to classic security vulnerabilities. It is designed to trace data flows throughout an application to detect potential leaks of sensitive information, such as Personally Identifiable Information (PII), API keys, and credentials. It supports multiple languages, including JavaScript and TypeScript, and aims to answer critical questions about how sensitive data is handled, stored, and transmitted.

\paragraph{Njsscan~\cite{njsscan}} This open-source tool is a static application security scanner specifically designed for Node.js applications. It leverages \textit{semgrep}~\cite{semgrep}, a powerful, syntax-aware semantic search tool, to run a ruleset tailored for the Node.js environment. This approach goes beyond simple regular expressions, allowing it to understand code structure to reduce false positives. Njsscan is representative of modern, community-driven SAST tools that provide targeted security analysis for a specific server-side JavaScript ecosystem.

\paragraph{JSPrime~\cite{repoJsprime}} Representing an earlier generation of JavaScript security tools, JSPrime is a foundational open-source tool known for its specific focus on client-side vulnerabilities. Its primary analysis technique was taint analysis, designed to trace data flows from user-controlled \textit{sources} to dangerous \textit{sinks} within the browser's Document Object Model (DOM). It was one of the first tools to effectively detect DOM-based Cross-Site Scripting (XSS) vulnerabilities. Including JSPrime provides a valuable historical baseline to compare the resilience of older taint-analysis engines with that of modern tools.

\paragraph{CodeQL~\cite{codeQl}} CodeQL, created by GitHub, converts a codebase into a relational database that can be queried using a specific declarative query language (QL). It does this by treating code as data. With the help of this paradigm, security researchers can create unique queries to find complex or specific vulnerability patterns that tools with preset detection logic might overlook. Complex data flows, control flows, and taint propagation paths throughout an entire application are all effectively modeled by CodeQL.

\paragraph{Snyk Code~\cite{synkCode}} Snyk Code, which is built for speed and ease of use and has deep integrations into developer workflows, source code repositories, and IDEs, embodies a modern, developer-first approach to application security. As developers write code, this enables real-time feedback, which aids in finding and fixing vulnerabilities. Its analysis engine stands out for using a semantic, artificial intelligence-based methodology that has been trained on a sizable collection of open-source code and carefully selected vulnerability disclosures. This enables the engine to recognize logical errors in the code by comprehending its context and intent, going beyond simple pattern matching.

\paragraph{ESLint with Security Plugins~\cite{eslintSecurityPlugin}} ESLint is widely used in the JavaScript ecosystem, despite being best known as a linter for detecting programmatic errors and enforcing code style. It can be made into a lightweight, pattern-matching SAST tool by adding security-specific plugins (like \textit{eslint-plugin-security}). Its analysis is mostly syntactic and relies on AST traversal to find known insecure patterns and anti-patterns, like hardcoded credentials, insecure regular expressions, and the use of \textit{eval()}.

We chose Njsscan and Bearer as the SAST tools to be assessed in this research study based on the previously established selection criteria. Although a number of other well-known tools were taken into consideration, they were eventually disregarded in order to guarantee a fair and methodologically consistent comparison. Despite being an effective and popular platform, SonarQube was not included because its free Community Edition only offers a small number of security guidelines in comparison to its paid versions. Importantly, paid tiers are the only ones that have the advanced taint-analysis rules required to identify a large number of injection-style vulnerabilities. This would have led to an unfair comparison with tools that have more extensive, publicly accessible feature sets. JSPrime was also excluded due to being technologically obsolete; its last update was nearly a decade ago, rendering it incapable of parsing modern JavaScript syntax. Despite our efforts to run it on our dataset, it frequently crashed and was unable to finish the analysis for the great majority of websites. We also did not include CodeQL. Despite being one of the most powerful SAST tools on the market, its analysis paradigm needs an extensive preprocessing step in which the source code is compiled into a database that can be queried. A straightforward and fair comparison of performance and scalability is unfair within the parameters of our study because of the significant differences between this customized, multi-step workflow and the other tools' direct-scan methodology. Lastly, we did not include ESLint. It is specifically made as a linter for finding local code smells and bugs, not for in-depth security analysis, even though its security plugins can identify some vulnerabilities. Furthermore, its analysis is limited to individual files and cannot record the data flows and dependencies between different JavaScript files, which means that its assessment of an application's security posture is fundamentally insufficient.

\section{Experimental Campaign}
\label{sec:experiments}

In this section, we present our experimental campaign, detailing how we applied the obfuscation configurations to the selected projects and systematically measured their impact on the vulnerability detection capabilities of the chosen SAST tools.

\subsection{Experimental Setup}
Our experimental campaign is designed to systematically quantify the impact of code obfuscation on the vulnerability detection capabilities of modern SAST tools. 
A fundamental requirement for this study was the complete availability of the source code; full access is strictly necessary to apply obfuscation techniques and to verify their impact against a verifiable baseline. To satisfy this requirement and conduct the study, we employed two distinct datasets. The first, a \textbf{Ground-Truth Dataset}, serves as a controlled benchmark composed of 16 deliberately vulnerable Node.js web applications. These projects were selected by including all web applications implemented with JavaScript technologies listed in the official OWASP Vulnerable Web Applications Directory~\cite{owaspVulnerableWebApps}. The second, an \textbf{In-the-Wild Dataset}, consists of 260 Node.js and JavaScript-based applications. To construct this dataset, we executed a targeted search query on GitHub designed to identify web applications built with these specific technologies. We collected the repositories resulting from this query, explicitly selecting freely available projects to ensure the code was fully open and accessible for our obfuscation experiments. This selection reflects a diversity of real-world coding practices while meeting the transparency requirements of our testing methodology. To better understand the differences between the two datasets, we compared them using standard software metrics (Table \ref{tab:metrics}), computed on each website with the Code Health Meter tool~\cite{codehealthmeter}.
The data shows a clear distinction in scale and complexity. The \textbf{In-the-Wild} projects are substantially larger, with an average size of roughly 14.800 lines of code (SLOC), which is nearly three times the average of the \textbf{Ground-Truth} applications (5.149). This gap is even more evident in the maximum values: the largest real-world project reaches nearly 250.000 lines of code, whereas the largest vulnerable-by-design application stops at around 42.500. Furthermore, the high standard deviation observed in the \textbf{In-the-Wild} dataset indicates a much wider variety of project structures. This confirms that while the \textbf{Ground-Truth} provides a controlled baseline, the \textbf{In-the-Wild} dataset is necessary to test our approach against the diverse and massive codebases found in actual production environments.

\begin{table}[!ht]
    \centering
    \small
    \setlength{\tabcolsep}{2pt}
    
    \begin{tabularx}{\columnwidth}{>{\raggedright\arraybackslash}X cccc} 
        \toprule
        \multicolumn{5}{c}{\textbf{Ground-Truth Dataset}} \\
        \midrule
        \textbf{Metric} & \textbf{Min} & \textbf{Avg} & \textbf{Max} & \textbf{Std. Dev.} \\
        \midrule
        Global Physical SLOC & 13.00 & 5149.25 & 42550 & 11403 \\
        CC per function Avg & 2.25 & 194.00 & 1449 & 402 \\
        Halstead Length per function & 51.19 & 2860.78 & 21081 & 5623 \\
        
        \midrule
        \multicolumn{5}{c}{\textbf{In-the-Wild Dataset}} \\
        \midrule
        \textbf{Metric} & \textbf{Min} & \textbf{Avg} & \textbf{Max} & \textbf{Std. Dev.} \\
        \midrule
        Global Physical SLOC & 2.00 & 14806.9 & 249769 & 33957 \\
        CC per function Avg & 1.00 & 50.34 & 2624 & 177 \\
        Halstead Length per function & 7.00 & 950.32 & 42448 & 3158 \\
        \bottomrule
    \end{tabularx}
    
    \vspace{5pt}
    \caption{Table of average metrics}
    \label{tab:metrics}
\end{table}

Finally, we assessed the vulnerability detection capabilities of two selected SAST tools, \textbf{Bearer} and \textbf{Njsscan}, across both datasets. To measure the reduction in tool efficacy, we define our primary metric, \textbf{Vulnerability Detection Loss (VDL)}. For any given project, we first establish a baseline by running the SAST tools on its original, unaltered version. We then apply one or more obfuscation techniques with javascript-obfuscator and run the tools again. The VDL quantifies the percentage of vulnerabilities, detected in the baseline, that are no longer reported by the tool on the obfuscated variant. Our methodology is structured to address the following research questions.

\paragraph{\textit{RQ1: What is the baseline performance of SAST tools on vulnerable-by-design websites, and how does this performance degrade by obfuscating their JavaScript code with one or multiple obfuscation techniques?}}
To establish a verifiable baseline, we first analyze each clean project in the \textbf{Ground-Truth Dataset} using Bearer and Njsscan, recording all detected (known) vulnerabilities. Subsequently, to assess the impact of layered obfuscation, we re-analyze each project after applying an increasing number of combined obfuscation techniques using the \textit{javascript-obfuscator} library. We then measure the VDL at each step to model the performance degradation curve.

\paragraph{\textit{RQ2: What is the individual impact of different obfuscation techniques on the vulnerability detection capabilities of SAST tools, when analysing real-world websites?}}
To isolate the effect of specific transformations, we apply each obfuscation method provided by \textit{javascript-obfuscator} in isolation to every project in our \textbf{In-the-Wild Dataset}. For each singly-obfuscated variant, we run the SAST tools and compute the resulting VDL. This approach allows us to rank the techniques based on their evasion efficacy on real-world websites and identify which transformations are most detrimental to SAST analysis.

\paragraph{\textit{RQ3: What is the cumulative effect of stacking multiple obfuscation techniques on the vulnerability detection capabilities of SAST tools, when analysing real-world websites?}}
To investigate this, we generate a comprehensive set of obfuscated variants by applying combinations of techniques simultaneously to every project in our \textbf{In-the-Wild Dataset}. We then categorize these variants by the number of active techniques. By analyzing the trend in VDL as the number of stacked techniques increases, we can determine the cumulative impact of obfuscation and identify any potential points of diminishing returns for the attacker.

\subsection{Benchmark Analysis}

To address RQ1, we evaluated the impact of obfuscation on the \textbf{Ground-Truth Dataset}, comprising 16 diverse Node.js projects. This analysis establishes the baseline performance degradation for each SAST tool when faced with known vulnerabilities. We applied a comprehensive set of obfuscation strategies, ranging from single techniques to stacked combinations, and computed the resulting VDL for each run.

\begin{figure}[!ht]
    \captionsetup[subfloat]{font=small, labelfont=scriptsize}
    
    \subfloat[\scriptsize Njsscan]{
        \begin{tikzpicture}
        \begin{axis}
        [
            cycle list={{black}},
            width=\columnwidth, 
            xmin=0,
            xmax=16.5,
            ymin=40,
            ymax=101,
            ytick={{40, 50, 60, 70, 80, 90, 100}},
            yticklabel={\pgfmathprintnumber{\tick}\%},
            yticklabel style={font=\footnotesize},
            boxplot/draw direction=y,
            ymajorgrids,
            xlabel={},
            xlabel style={yshift=-20pt},
            ylabel={Overall Vulnerability Loss (\%)},
            ylabel style={yshift=-5pt},
            xtick={1, 2, 3, 4, 5, 6, 7, 8, 9, 10, 11, 12, 13, 14, 15, 16},
            xticklabels={DVNA, DVWS, Damn-Vuln-GraphQL, NodeGoat, Vulnerable-nodejs, VulnerableApp, VulnerableApp-facade, Zero-Health, appsecco-dvna, chat.js, juice-shop, vuln-nodejs-app, vulnerable-node, websheep, wrongsecrets, ypreyAPINodeJS},
            xticklabel style={rotate=45, anchor=east, font=\scriptsize}
        ]

        \addplot+ [
            boxplot prepared={
                draw position=1,
                median=100.0,
                upper quartile=100.0,
                lower quartile=100.0,
                upper whisker=100.0,
                lower whisker=100.0
            },
            fill=cyan!40, draw=black,
        ] coordinates {}; 

        \addplot+ [
            boxplot prepared={
                draw position=3,
                median=100.0,
                upper quartile=100.0,
                lower quartile=100.0,
                upper whisker=100.0,
                lower whisker=100.0
            },
            fill=cyan!40, draw=black,
        ] coordinates {}; 

        \addplot+ [
            boxplot prepared={
                draw position=4,
                median=90.9090909090909,
                upper quartile=100.0,
                lower quartile=81.81818181818183,
                upper whisker=100.0,
                lower whisker=63.63636363636363
            },
            fill=cyan!40, draw=black,
        ] coordinates {}; 

        \addplot+ [
            boxplot prepared={
                draw position=5,
                median=100.0,
                upper quartile=100.0,
                lower quartile=100.0,
                upper whisker=100.0,
                lower whisker=100.0
            },
            fill=cyan!40, draw=black,
        ] coordinates {}; 

        \addplot [only marks, mark=*, mark size=1.5pt] coordinates {(5,60.0) (5,60.0) (5,60.0) (5,60.0) (5,60.0) (5,60.0) (5,60.0) (5,60.0) (5,60.0) (5,60.0) (5,60.0) (5,60.0) (5,60.0) (5,60.0) (5,60.0) (5,60.0) (5,60.0) (5,60.0) (5,60.0) (5,60.0) (5,60.0) (5,60.0) (5,60.0) (5,60.0) (5,60.0) (5,60.0) (5,60.0) (5,60.0) (5,60.0) (5,60.0) (5,60.0) (5,60.0) (5,60.0) (5,60.0) (5,60.0) (5,60.0) (5,60.0) (5,60.0) (5,60.0) (5,60.0) (5,60.0) (5,60.0) (5,60.0) (5,60.0) (5,60.0) (5,60.0) (5,60.0) (5,60.0) (5,60.0) (5,60.0) (5,60.0) (5,60.0) (5,60.0) (5,60.0) (5,60.0) (5,60.0) (5,60.0) (5,60.0) (5,60.0) (5,60.0) (5,60.0)};

        \addplot+ [
            boxplot prepared={
                draw position=8,
                median=100.0,
                upper quartile=100.0,
                lower quartile=85.71428571428571,
                upper whisker=100.0,
                lower whisker=85.71428571428571
            },
            fill=cyan!40, draw=black,
        ] coordinates {}; 

        \addplot+ [
            boxplot prepared={
                draw position=9,
                median=75.0,
                upper quartile=87.5,
                lower quartile=75.0,
                upper whisker=87.5,
                lower whisker=62.5
            },
            fill=cyan!40, draw=black,
        ] coordinates {}; 

        \addplot [only marks, mark=*, mark size=1.5pt] coordinates {(9,50.0) (9,50.0) (9,50.0) (9,50.0) (9,50.0) (9,50.0) (9,50.0) (9,50.0) (9,50.0) (9,50.0) (9,50.0) (9,50.0) (9,50.0) (9,50.0) (9,50.0) (9,50.0) (9,50.0) (9,50.0) (9,50.0) (9,50.0) (9,50.0) (9,50.0) (9,50.0) (9,50.0) (9,50.0) (9,50.0) (9,50.0) (9,50.0) (9,50.0) (9,50.0) (9,50.0) (9,50.0) (9,50.0) (9,50.0) (9,50.0) (9,50.0)};

        \addplot+ [
            boxplot prepared={
                draw position=10,
                median=100.0,
                upper quartile=100.0,
                lower quartile=100.0,
                upper whisker=100.0,
                lower whisker=100.0
            },
            fill=cyan!40, draw=black,
        ] coordinates {}; 

        \addplot [only marks, mark=*, mark size=1.5pt] coordinates {(10,66.66666666666666) (10,66.66666666666666) (10,66.66666666666666) (10,66.66666666666666) (10,66.66666666666666) (10,66.66666666666666) (10,66.66666666666666) (10,66.66666666666666) (10,66.66666666666666) (10,66.66666666666666) (10,66.66666666666666) (10,66.66666666666666) (10,66.66666666666666) (10,66.66666666666666) (10,66.66666666666666) (10,66.66666666666666) (10,66.66666666666666) (10,66.66666666666666) (10,66.66666666666666) (10,66.66666666666666) (10,66.66666666666666) (10,66.66666666666666) (10,66.66666666666666) (10,66.66666666666666) (10,66.66666666666666) (10,66.66666666666666) (10,66.66666666666666) (10,66.66666666666666) (10,66.66666666666666) (10,66.66666666666666) (10,66.66666666666666) (10,66.66666666666666) (10,66.66666666666666) (10,66.66666666666666) (10,66.66666666666666) (10,66.66666666666666) (10,66.66666666666666) (10,66.66666666666666) (10,66.66666666666666) (10,66.66666666666666) (10,66.66666666666666) (10,66.66666666666666) (10,66.66666666666666) (10,66.66666666666666) (10,66.66666666666666) (10,66.66666666666666) (10,66.66666666666666) (10,66.66666666666666) (10,66.66666666666666) (10,66.66666666666666) (10,66.66666666666666) (10,66.66666666666666) (10,66.66666666666666) (10,66.66666666666666) (10,66.66666666666666) (10,66.66666666666666)};

        \addplot+ [
            boxplot prepared={
                draw position=11,
                median=100.0,
                upper quartile=100.0,
                lower quartile=100.0,
                upper whisker=100.0,
                lower whisker=100.0
            },
            fill=cyan!40, draw=black,
        ] coordinates {}; 

        \addplot+ [
            boxplot prepared={
                draw position=13,
                median=83.33333333333334,
                upper quartile=83.33333333333334,
                lower quartile=83.33333333333334,
                upper whisker=83.33333333333334,
                lower whisker=83.33333333333334
            },
            fill=cyan!40, draw=black,
        ] coordinates {}; 

        \addplot [only marks, mark=*, mark size=1.5pt] coordinates {(13,50.0) (13,50.0) (13,50.0) (13,50.0) (13,50.0) (13,50.0) (13,50.0) (13,50.0) (13,50.0) (13,50.0) (13,50.0) (13,50.0) (13,50.0) (13,50.0) (13,50.0) (13,50.0) (13,50.0) (13,50.0) (13,50.0) (13,50.0) (13,50.0) (13,50.0) (13,50.0) (13,50.0) (13,50.0) (13,50.0) (13,50.0) (13,50.0) (13,50.0) (13,50.0) (13,50.0) (13,50.0) (13,50.0) (13,50.0) (13,50.0) (13,50.0) (13,50.0) (13,50.0) (13,50.0) (13,50.0) (13,50.0) (13,50.0) (13,50.0) (13,50.0) (13,50.0) (13,50.0) (13,50.0) (13,50.0) (13,50.0) (13,50.0) (13,50.0) (13,50.0) (13,50.0) (13,50.0)};

        \addplot+ [
            boxplot prepared={
                draw position=15,
                median=100.0,
                upper quartile=100.0,
                lower quartile=100.0,
                upper whisker=100.0,
                lower whisker=100.0
            },
            fill=cyan!40, draw=black,
        ] coordinates {}; 

    \end{axis}
        \end{tikzpicture}
        \label{fig:vdl_benchmark_njsscan}
    }
    
    \subfloat[\scriptsize Bearer]{
    \begin{tikzpicture}
    \begin{axis}
    [
        cycle list={{black}},
        width=\columnwidth,
        xmin=0,
        xmax=16.5,
        boxplot/draw direction=y,
        ymajorgrids,
        ymin=-10,
        ymax=101,
        ylabel={Vulnerability Loss (\%)},
        ylabel style={yshift=-5pt},
        yticklabel={\pgfmathprintnumber{\tick}\%},
        yticklabel style={font=\footnotesize},
        xtick={1, 2, 3, 4, 5, 6, 7, 8, 9, 10, 11, 12, 13, 14, 15, 16},
        xticklabels={DVNA, DVWS, Damn-Vuln-GraphQL, NodeGoat, Vulnerable-nodejs, VulnerableApp, VulnerableApp-facade, Zero-Health, appsecco-dvna, chat.js, juice-shop, vuln-nodejs-app, vulnerable-node, websheep, wrongsecrets, ypreyAPINodeJS},
        xticklabel style={rotate=45, anchor=east, font=\scriptsize}
    ]

        \addplot+ [
            boxplot prepared={
                draw position=1,
                median=80.0,
                upper quartile=84.0,
                lower quartile=68.0,
                upper whisker=100.0,
                lower whisker=60.0
            },
            fill=cyan!40, draw=black,
            boxplot/median/.style={draw=firebrick, thick, line cap=rect},
        ] coordinates {}; 

        \addplot+ [
            boxplot prepared={
                draw position=4,
                median=92.45283018867924,
                upper quartile=98.11320754716981,
                lower quartile=86.79245283018868,
                upper whisker=100.0,
                lower whisker=75.47169811320755
            },
            fill=cyan!40, draw=black,
            boxplot/median/.style={draw=firebrick, thick, line cap=rect},
        ] coordinates {}; 

        \addplot+ [
            boxplot prepared={
                draw position=5,
                median=75.0,
                upper quartile=100.0,
                lower quartile=66.66666666666666,
                upper whisker=100.0,
                lower whisker=66.66666666666666
            },
            fill=cyan!40, draw=black,
            boxplot/median/.style={draw=firebrick, thick, line cap=rect},
        ] coordinates {}; 

        \addplot+ [
            boxplot prepared={
                draw position=6,
                median=100.0,
                upper quartile=100.0,
                lower quartile=100.0,
                upper whisker=100.0,
                lower whisker=100.0
            },
            fill=cyan!40, draw=black,
            boxplot/median/.style={draw=firebrick, thick, line cap=rect},
        ] coordinates {}; 

        \addplot+ [
            boxplot prepared={
                draw position=7,
                median=0.0,
                upper quartile=0.0,
                lower quartile=0.0,
                upper whisker=0.0,
                lower whisker=0.0
            },
            fill=cyan!40, draw=black,
            boxplot/median/.style={draw=firebrick, thick, line cap=rect},
        ] coordinates {}; 

        \addplot+ [
            boxplot prepared={
                draw position=8,
                median=87.78625954198473,
                upper quartile=87.78625954198473,
                lower quartile=87.02290076335878,
                upper whisker=87.78625954198473,
                lower whisker=86.25954198473282
            },
            fill=cyan!40, draw=black,
            boxplot/median/.style={draw=firebrick, thick, line cap=rect},
        ] coordinates {}; 

        \addplot [only marks, mark=*, mark size=1.5pt] coordinates {(8,85.49618320610686) (8,83.96946564885496) (8,83.20610687022901) (8,84.7328244274809) (8,83.96946564885496) (8,84.7328244274809) (8,83.20610687022901) (8,82.44274809160305) (8,84.7328244274809) (8,84.7328244274809) (8,84.7328244274809) (8,83.20610687022901) (8,83.20610687022901) (8,84.7328244274809) (8,83.96946564885496) (8,85.49618320610686) (8,85.49618320610686) (8,85.49618320610686) (8,84.7328244274809) (8,83.20610687022901) (8,83.20610687022901) (8,83.20610687022901)};

        \addplot+ [
            boxplot prepared={
                draw position=9,
                median=90.32258064516128,
                upper quartile=100.0,
                lower quartile=87.09677419354838,
                upper whisker=100.0,
                lower whisker=80.64516129032258
            },
            fill=cyan!40, draw=black,
            boxplot/median/.style={draw=firebrick, thick, line cap=rect},
        ] coordinates {}; 

        \addplot+ [
            boxplot prepared={
                draw position=10,
                median=100.0,
                upper quartile=100.0,
                lower quartile=100.0,
                upper whisker=100.0,
                lower whisker=100.0
            },
            fill=cyan!40, draw=black,
            boxplot/median/.style={draw=firebrick, thick, line cap=rect},
        ] coordinates {}; 

        \addplot [only marks, mark=*, mark size=1.5pt] coordinates {(10,80.0) (10,80.0)};

        \addplot+ [
            boxplot prepared={
                draw position=11,
                median=0.2915451895043732,
                upper quartile=0.2915451895043732,
                lower quartile=0.2915451895043732,
                upper whisker=0.2915451895043732,
                lower whisker=0.2915451895043732
            },
            fill=cyan!40, draw=black,
            boxplot/median/.style={draw=firebrick, thick, line cap=rect},
        ] coordinates {}; 

        \addplot+ [
            boxplot prepared={
                draw position=13,
                median=91.66666666666666,
                upper quartile=95.83333333333334,
                lower quartile=79.16666666666666,
                upper whisker=95.83333333333334,
                lower whisker=70.83333333333334
            },
            fill=cyan!40, draw=black,
            boxplot/median/.style={draw=firebrick, thick, line cap=rect},
        ] coordinates {}; 

        \addplot+ [
            boxplot prepared={
                draw position=14,
                median=0.0,
                upper quartile=0.0,
                lower quartile=0.0,
                upper whisker=0.0,
                lower whisker=0.0
            },
            fill=cyan!40, draw=black,
            boxplot/median/.style={draw=firebrick, thick, line cap=rect},
        ] coordinates {}; 

        \addplot+ [
            boxplot prepared={
                draw position=16,
                median=100.0,
                upper quartile=100.0,
                lower quartile=100.0,
                upper whisker=100.0,
                lower whisker=100.0
            },
            fill=cyan!40, draw=black,
            boxplot/median/.style={draw=firebrick, thick, line cap=rect},
        ] coordinates {}; 

    \end{axis}
    \end{tikzpicture}
    \label{fig:vdl_benchmark_bearer}
    }
    \refstepcounter{figure}
    \caption{VDL distribution for Bearer and Njsscan on the benchmark dataset. Each plot aggregates VDL results from all single and stacked obfuscation combinations.}
    \label{fig:vdl_benchmark}
\end{figure}
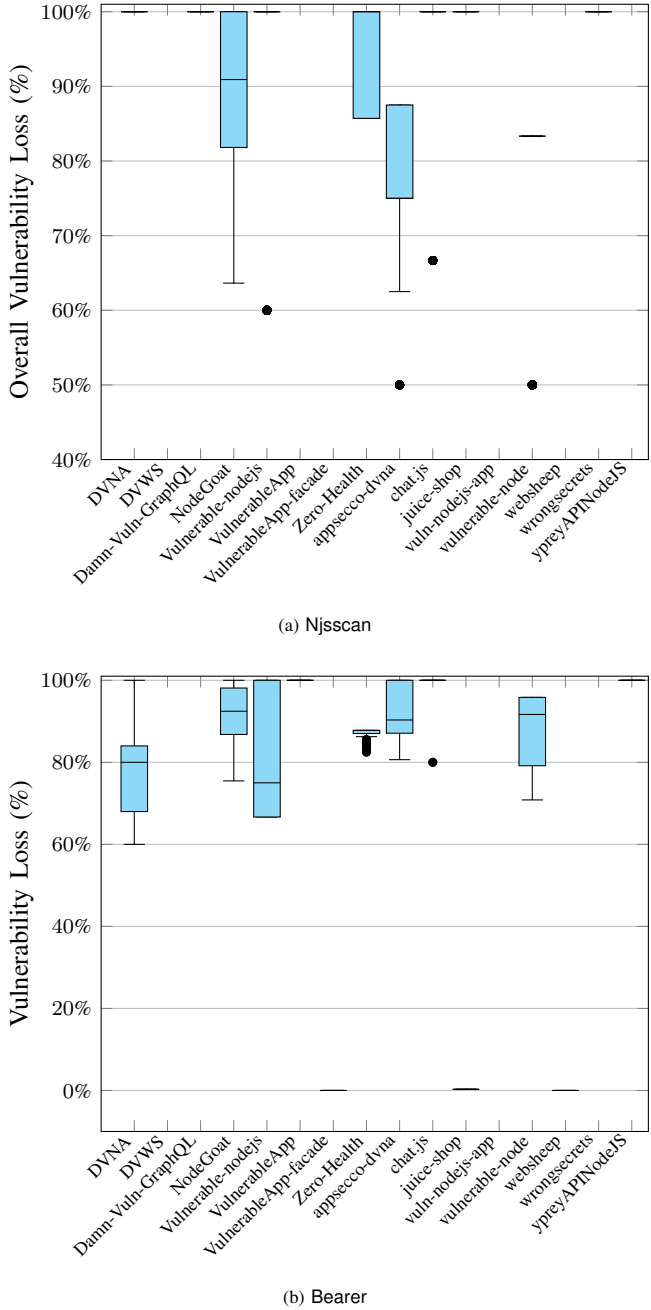 
Figure~\ref{fig:vdl_benchmark} illustrates the distribution of VDL for each benchmark project, aggregated across all applied obfuscation combinations. The results are presented separately for Njsscan (~\ref{fig:vdl_benchmark_njsscan}) and Bearer (~\ref{fig:vdl_benchmark_bearer}).

\paragraph{Njsscan (Figure~\ref{fig:vdl_benchmark_njsscan})}
The analysis for Njsscan reveals significant limitations in both baseline detection and resilience.
First, a substantial subset of projects, specifically \textit{DVWS}, \textit{VulnerableApp}, \textit{VulnerableApp-facade}, \textit{vuln-nodejs-app}, \textit{websheep}, and \textit{ypreyAPINodeJS}, exhibit empty plots in the figure. This absence of data indicates that Njsscan failed to detect any vulnerabilities in these projects even in their original, non-obfuscated state (Zero Baseline Detection). Consequently, VDL could not be computed.

For the remaining projects where baseline vulnerabilities were detected, the tool exhibits high susceptibility to obfuscation. Projects such as \textit{DVNA}, \textit{Damn-Vulnerable-GraphQL}, \textit{Vulnerable-nodejs}, \textit{Zero-Health}, \textit{chat.js}, \textit{juice-shop}, and \textit{wrongsecrets} show a median VDL of 100\%. In these cases, the obfuscation successfully masks every single vulnerability. Only a few intermediate cases, such as \textit{appsecco-dvna} (median $\approx$75\%) and \textit{vulnerable-node} (median $\approx$83\%), show slight resistance, though the loss remains critically high. This confirms that Njsscan's regex-based static analysis is highly fragile against code transformations.

\paragraph{Bearer (Figure~\ref{fig:vdl_benchmark_bearer})}
Bearer demonstrates a more varied performance profile but shares some baseline limitations. Similar to Njsscan, strictly empty plots are observed for \textit{DVWS}, \textit{Damn-Vulnerable-GraphQL}, \textit{vuln-nodejs-app}, and \textit{wrongsecrets}, indicating that Bearer also failed to identify any baseline vulnerabilities in these specific architectures.

However, unlike Njsscan, Bearer displays instances of remarkable resilience. It maintains near-perfect detection (median VDL $\approx$0\%) on \textit{VulnerableApp-facade}, \textit{websheep}, and \textit{juice-shop}, suggesting that its analysis engine can effectively penetrate obfuscation in these specific codebases.
Conversely, it suffers total collapse (100\% VDL) on \textit{VulnerableApp}, \textit{chat.js}, and \textit{ypreyAPINodeJS}. The remaining projects (e.g., \textit{NodeGoat}, \textit{DVNA}) generally exhibit high vulnerability loss (ranging from 75\% to 90\%), indicating that while Bearer is slightly more robust than Njsscan in specific instances, it still struggles significantly with obfuscated Node.js code.

\begin{tcolorbox}[colback=gray!5,colframe=blue!40, boxrule=0.3mm]
\textbf{RQ1:} Baseline capabilities significantly impact the evaluation. A notable portion of the dataset resulted in empty plots (e.g., \textit{DVWS}, \textit{vuln-nodejs-app}) for both tools, signifying a failure to detect vulnerabilities even before obfuscation. Where detection occurred, \textbf{Njsscan} proved extremely fragile, with median VDLs hitting 100\% on most projects. \textbf{Bearer} showed higher variance, achieving high resilience (0\% VDL) on three projects but succumbing to high loss on the majority of the benchmark. [See Figure~\ref{fig:vdl_benchmark}]
\end{tcolorbox}

\subsection{In-The-Wild Analysis}

The severe performance degradation observed on the benchmark dataset (RQ1) needs an investigation into whether these results generalize to a larger, more heterogeneous corpus. We thus proceed to analyze the \textbf{In-the-Wild Dataset}. This phase enables us to isolate the impact of obfuscation techniques, both in isolation (addressing RQ2) and in combination (addressing RQ3).

The \textbf{In-the-Wild Dataset} includes 290 websites gathered from GitHub. We report that Bearer and njsscan reported no vulnerabilities on 30 out of 290 websites included in this dataset. Thus, we removed such (apparently) not vulnerable websites from the \textbf{In-the-Wild Dataset} before generating the obfuscated variants.

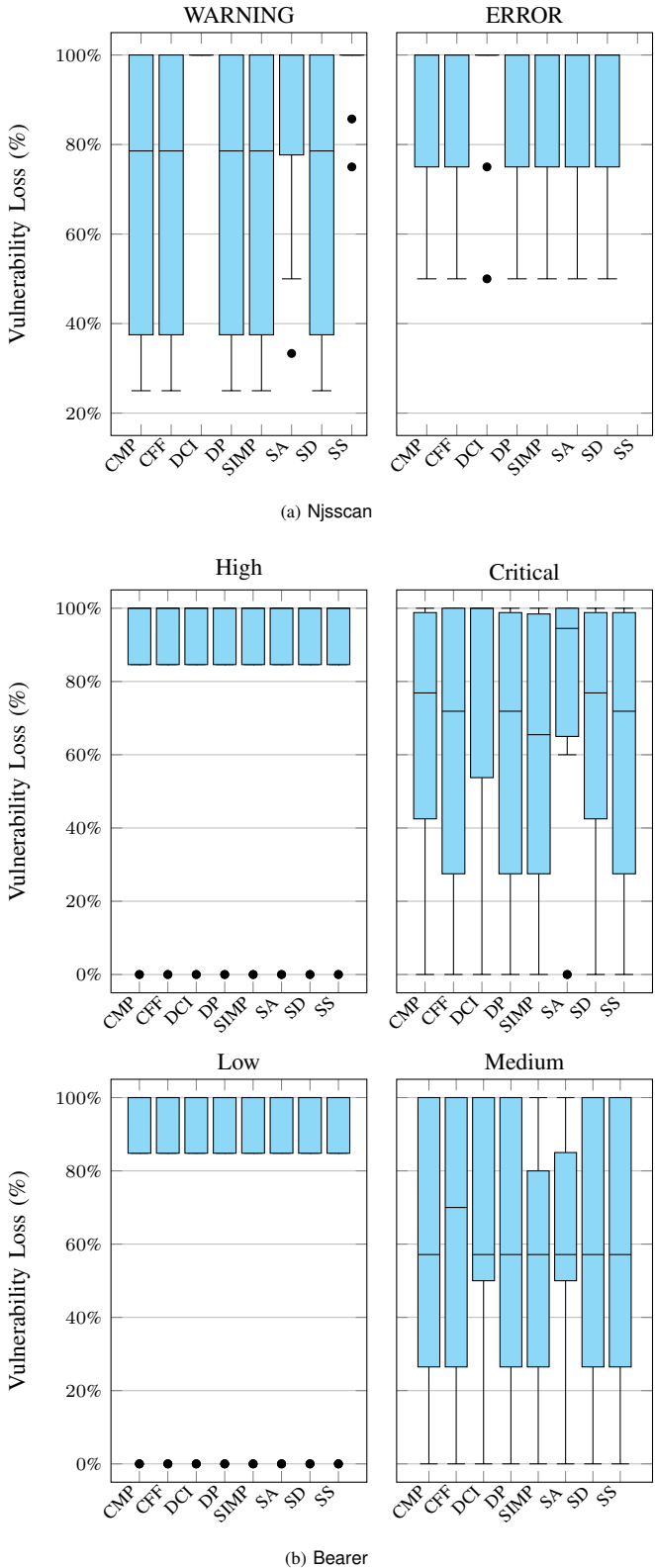
\begin{figure}[!ht]
    \captionsetup[subfloat]{font=small, labelfont=scriptsize}
    \centering
    \subfloat[\scriptsize Njsscan]{
        \begin{tikzpicture}
        \begin{groupplot}[
            cycle list={{black}},
            group style={
                group size=2 by 1,
                horizontal sep=0.4cm
            },
            width=5.02cm,
            height=7.0cm,
            ymin=15,
            ymax=105,
            ymajorgrids,
            boxplot/draw direction=y,
            ylabel={},
            yticklabel={},
            yticklabel={\pgfmathprintnumber{\tick}\%},
            tick label style={font=\scriptsize},
            label style={font=\small},
            title style={font=\small, yshift=-0.2cm}
        ]
    
        \nextgroupplot[
            cycle list={{black}},
            title={WARNING},
            xlabel={},
            xmin=0,
            xmax=8.5,
            xtick={1, 2, 3, 4, 5, 6, 7, 8},
            xticklabels={CMP, CFF, DCI, DP, SIMP, SA, SD, SS},
            xticklabel style={rotate=45, anchor=east},
            ytick={20, 40, 60, 80, 100},
            yticklabel={\pgfmathprintnumber{\tick}\%},
            ylabel={Vulnerability Loss (\%)}
        ]
    
            \addplot+ [
                boxplot prepared={
                    median=78.57142857142857,
                    upper quartile=100.0,
                    lower quartile=37.5,
                    upper whisker=100.0,
                    lower whisker=25.0
                },
                fill=cyan!40, draw=black,
            ] coordinates {};
    
            \addplot+ [
                boxplot prepared={
                    median=78.57142857142857,
                    upper quartile=100.0,
                    lower quartile=37.5,
                    upper whisker=100.0,
                    lower whisker=25.0
                },
                fill=cyan!40, draw=black,
            ] coordinates {};
    
            \addplot+ [
                boxplot prepared={
                    median=100.0,
                    upper quartile=100.0,
                    lower quartile=100.0,
                    upper whisker=100.0,
                    lower whisker=100.0
                },
                fill=cyan!40, draw=black,
            ] coordinates {};
    
            \addplot+ [
                boxplot prepared={
                    median=78.57142857142857,
                    upper quartile=100.0,
                    lower quartile=37.5,
                    upper whisker=100.0,
                    lower whisker=25.0
                },
                fill=cyan!40, draw=black,
            ] coordinates {};
    
            \addplot+ [
                boxplot prepared={
                    median=78.57142857142857,
                    upper quartile=100.0,
                    lower quartile=37.5,
                    upper whisker=100.0,
                    lower whisker=25.0
                },
                fill=cyan!40, draw=black,
            ] coordinates {};
    
            \addplot+ [
                boxplot prepared={
                    median=100.0,
                    upper quartile=100.0,
                    lower quartile=77.67857142857143,
                    upper whisker=100.0,
                    lower whisker=50.0
                },
                fill=cyan!40, draw=black,
            ] coordinates {};
    
            \addplot [only marks, mark=*, mark size=1.5pt] coordinates {(6,33.33333333333333)};
    
            \addplot+ [
                boxplot prepared={
                    median=78.57142857142857,
                    upper quartile=100.0,
                    lower quartile=37.5,
                    upper whisker=100.0,
                    lower whisker=25.0
                },
                fill=cyan!40, draw=black,
            ] coordinates {};
    
            \addplot+ [
                boxplot prepared={
                    median=100.0,
                    upper quartile=100.0,
                    lower quartile=100.0,
                    upper whisker=100.0,
                    lower whisker=100.0
                },
                fill=cyan!40, draw=black,
            ] coordinates {};
    
            \addplot [only marks, mark=*, mark size=1.5pt] coordinates {(8,85.71428571428571) (8,75.0)};
    
        \nextgroupplot[
            cycle list={{black}},
            title={ERROR},
            xmin=0,
            xmax=8.5,
            xlabel={},
            xtick={1, 2, 3, 4, 5, 6, 7, 8},
            xticklabels={CMP, CFF, DCI, DP, SIMP, SA, SD, SS},
            xticklabel style={rotate=45, anchor=east},
            ytick={20, 40, 60, 80, 100},
            yticklabel=\empty
        ]
    
            \addplot+ [
                boxplot prepared={
                    median=100.0,
                    upper quartile=100.0,
                    lower quartile=75.0,
                    upper whisker=100.0,
                    lower whisker=50.0
                },
                fill=cyan!40, draw=black,
            ] coordinates {};
    
            \addplot+ [
                boxplot prepared={
                    median=100.0,
                    upper quartile=100.0,
                    lower quartile=75.0,
                    upper whisker=100.0,
                    lower whisker=50.0
                },
                fill=cyan!40, draw=black,
            ] coordinates {};
    
            \addplot+ [
                boxplot prepared={
                    median=100.0,
                    upper quartile=100.0,
                    lower quartile=100.0,
                    upper whisker=100.0,
                    lower whisker=100.0
                },
                fill=cyan!40, draw=black,
            ] coordinates {};
    
            \addplot [only marks, mark=*, mark size=1.5pt] coordinates {(3,50.0) (3,75.0)};
    
            \addplot+ [
                boxplot prepared={
                    median=100.0,
                    upper quartile=100.0,
                    lower quartile=75.0,
                    upper whisker=100.0,
                    lower whisker=50.0
                },
                fill=cyan!40, draw=black,
            ] coordinates {};
    
            \addplot+ [
                boxplot prepared={
                    median=100.0,
                    upper quartile=100.0,
                    lower quartile=75.0,
                    upper whisker=100.0,
                    lower whisker=50.0
                },
                fill=cyan!40, draw=black,
            ] coordinates {};
    
            \addplot+ [
                boxplot prepared={
                    median=100.0,
                    upper quartile=100.0,
                    lower quartile=75.0,
                    upper whisker=100.0,
                    lower whisker=50.0
                },
                fill=cyan!40, draw=black,
            ] coordinates {};
    
            \addplot+ [
                boxplot prepared={
                    median=100.0,
                    upper quartile=100.0,
                    lower quartile=75.0,
                    upper whisker=100.0,
                    lower whisker=50.0
                },
                fill=cyan!40, draw=black,
            ] coordinates {};
    
        \end{groupplot}
        \end{tikzpicture}
        \label{fig:vdl_single_plugin_in_the_wild_njsscan}
    }

\subfloat[\scriptsize Bearer]{
        \begin{tikzpicture}
        \begin{groupplot}[
            cycle list={{black}},
            group style={
                group size=2 by 2,
                vertical sep=-12.0cm,
                horizontal sep=0.4cm
            },
            width=5.02cm,
            height=7.0cm,
            ymin=-5,
            ymax=105,
            ymajorgrids,
            boxplot/draw direction=y,
            ylabel={},
            yticklabel={},
            yticklabel={\pgfmathprintnumber{\tick}\%},
            tick label style={font=\scriptsize},
            label style={font=\small},
            title style={font=\small, yshift=-0.2cm}
        ]
    
        \nextgroupplot[
            title={Low},
            xmin=0,
            xmax=9,
            xlabel={},
            xtick={1, 2, 3, 4, 5, 6, 7, 8},
            ylabel={Vulnerability Loss (\%)},
            xticklabels={CMP, CFF, DCI, DP, SIMP, SA, SD, SS},
            xticklabel style={rotate=45, anchor=east}
        ]
    
            \addplot+ [
                boxplot prepared={
                    median=100.0,
                    upper quartile=100.0,
                    lower quartile=84.81012658227847,
                    upper whisker=100.0,
                    lower whisker=84.81012658227847
                },
                fill=cyan!40, draw=black,
            ] coordinates {};
    
            \addplot [only marks, mark=*, mark size=1.5pt] coordinates {(1,0.0) (1,0.0) (1,0.0)};
    
            \addplot+ [
                boxplot prepared={
                    median=100.0,
                    upper quartile=100.0,
                    lower quartile=84.81012658227847,
                    upper whisker=100.0,
                    lower whisker=84.81012658227847
                },
                fill=cyan!40, draw=black,
            ] coordinates {};
    
            \addplot [only marks, mark=*, mark size=1.5pt] coordinates {(2,0.0) (2,0.0) (2,0.0)};
    
            \addplot+ [
                boxplot prepared={
                    median=100.0,
                    upper quartile=100.0,
                    lower quartile=84.81012658227847,
                    upper whisker=100.0,
                    lower whisker=84.81012658227847
                },
                fill=cyan!40, draw=black,
            ] coordinates {};
    
            \addplot [only marks, mark=*, mark size=1.5pt] coordinates {(3,0.0) (3,0.0) (3,0.0)};
    
            \addplot+ [
                boxplot prepared={
                    median=100.0,
                    upper quartile=100.0,
                    lower quartile=84.81012658227847,
                    upper whisker=100.0,
                    lower whisker=84.81012658227847
                },
                fill=cyan!40, draw=black,
            ] coordinates {};
    
            \addplot [only marks, mark=*, mark size=1.5pt] coordinates {(4,0.0) (4,0.0) (4,0.0)};
    
            \addplot+ [
                boxplot prepared={
                    median=100.0,
                    upper quartile=100.0,
                    lower quartile=84.81012658227847,
                    upper whisker=100.0,
                    lower whisker=84.81012658227847
                },
                fill=cyan!40, draw=black,
            ] coordinates {};
    
            \addplot [only marks, mark=*, mark size=1.5pt] coordinates {(5,0.0) (5,0.0) (5,0.0)};
    
            \addplot+ [
                boxplot prepared={
                    median=100.0,
                    upper quartile=100.0,
                    lower quartile=84.81012658227847,
                    upper whisker=100.0,
                    lower whisker=84.81012658227847
                },
                fill=cyan!40, draw=black,
            ] coordinates {};
    
            \addplot [only marks, mark=*, mark size=1.5pt] coordinates {(6,0.0) (6,0.0) (6,0.0)};
    
            \addplot+ [
                boxplot prepared={
                    median=100.0,
                    upper quartile=100.0,
                    lower quartile=84.81012658227847,
                    upper whisker=100.0,
                    lower whisker=84.81012658227847
                },
                fill=cyan!40, draw=black,
            ] coordinates {};
    
            \addplot [only marks, mark=*, mark size=1.5pt] coordinates {(7,0.0) (7,0.0) (7,0.0)};
    
            \addplot+ [
                boxplot prepared={
                    median=100.0,
                    upper quartile=100.0,
                    lower quartile=84.81012658227847,
                    upper whisker=100.0,
                    lower whisker=84.81012658227847
                },
                fill=cyan!40, draw=black,
            ] coordinates {};
    
            \addplot [only marks, mark=*, mark size=1.5pt] coordinates {(8,0.0) (8,0.0) (8,0.0)};
    
        \nextgroupplot[
            cycle list={{black}},
            title={Medium},
            xlabel={},
            xtick={1, 2, 3, 4, 5, 6, 7, 8},
            xticklabels={CMP, CFF, DCI, DP, SIMP, SA, SD, SS},
            xticklabel style={rotate=45, anchor=east},
            yticklabels={}
        ]
    
            \addplot+ [
                boxplot prepared={
                    median=57.14285714285714,
                    upper quartile=100.0,
                    lower quartile=26.470588235294116,
                    upper whisker=100.0,
                    lower whisker=0.0
                },
                fill=cyan!40, draw=black,
            ] coordinates {};
    
            \addplot+ [
                boxplot prepared={
                    median=70.0,
                    upper quartile=100.0,
                    lower quartile=26.470588235294116,
                    upper whisker=100.0,
                    lower whisker=0.0
                },
                fill=cyan!40, draw=black,
            ] coordinates {};
    
            \addplot+ [
                boxplot prepared={
                    median=57.14285714285714,
                    upper quartile=100.0,
                    lower quartile=50.0,
                    upper whisker=100.0,
                    lower whisker=0.0
                },
                fill=cyan!40, draw=black,
            ] coordinates {};
    
            \addplot+ [
                boxplot prepared={
                    median=57.14285714285714,
                    upper quartile=100.0,
                    lower quartile=26.470588235294116,
                    upper whisker=100.0,
                    lower whisker=0.0
                },
                fill=cyan!40, draw=black,
            ] coordinates {};
    
            \addplot+ [
                boxplot prepared={
                    median=57.14285714285714,
                    upper quartile=80.0,
                    lower quartile=26.470588235294116,
                    upper whisker=100.0,
                    lower whisker=0.0
                },
                fill=cyan!40, draw=black,
            ] coordinates {};
    
            \addplot+ [
                boxplot prepared={
                    median=57.14285714285714,
                    upper quartile=85.0,
                    lower quartile=50.0,
                    upper whisker=100.0,
                    lower whisker=0.0
                },
                fill=cyan!40, draw=black,
            ] coordinates {};
    
            \addplot+ [
                boxplot prepared={
                    median=57.14285714285714,
                    upper quartile=100.0,
                    lower quartile=26.470588235294116,
                    upper whisker=100.0,
                    lower whisker=0.0
                },
                fill=cyan!40, draw=black,
            ] coordinates {};
    
            \addplot+ [
                boxplot prepared={
                    median=57.14285714285714,
                    upper quartile=100.0,
                    lower quartile=26.470588235294116,
                    upper whisker=100.0,
                    lower whisker=0.0
                },
                fill=cyan!40, draw=black,
            ] coordinates {};
    
        \nextgroupplot[
            cycle list={{black}},
            title={High},
            xmin=0,
            xmax=9,
            xlabel={},
            xtick={1, 2, 3, 4, 5, 6, 7, 8},
            ylabel={Vulnerability Loss (\%)},
            xticklabels={CMP, CFF, DCI, DP, SIMP, SA, SD, SS},
            xticklabel style={rotate=45, anchor=east}
        ]
    
            \addplot+ [
                boxplot prepared={
                    median=100.0,
                    upper quartile=100.0,
                    lower quartile=84.61538461538461,
                    upper whisker=100.0,
                    lower whisker=84.61538461538461
                },
                fill=cyan!40, draw=black,
            ] coordinates {};
    
            \addplot [only marks, mark=*, mark size=1.5pt] coordinates {(1,0.0) (1,0.0)};
    
            \addplot+ [
                boxplot prepared={
                    median=100.0,
                    upper quartile=100.0,
                    lower quartile=84.61538461538461,
                    upper whisker=100.0,
                    lower whisker=84.61538461538461
                },
                fill=cyan!40, draw=black,
            ] coordinates {};
    
            \addplot [only marks, mark=*, mark size=1.5pt] coordinates {(2,0.0) (2,0.0)};
    
            \addplot+ [
                boxplot prepared={
                    median=100.0,
                    upper quartile=100.0,
                    lower quartile=84.61538461538461,
                    upper whisker=100.0,
                    lower whisker=84.61538461538461
                },
                fill=cyan!40, draw=black,
            ] coordinates {};
    
            \addplot [only marks, mark=*, mark size=1.5pt] coordinates {(3,0.0) (3,0.0)};
    
            \addplot+ [
                boxplot prepared={
                    median=100.0,
                    upper quartile=100.0,
                    lower quartile=84.61538461538461,
                    upper whisker=100.0,
                    lower whisker=84.61538461538461
                },
                fill=cyan!40, draw=black,
            ] coordinates {};
    
            \addplot [only marks, mark=*, mark size=1.5pt] coordinates {(4,0.0) (4,0.0)};
    
            \addplot+ [
                boxplot prepared={
                    median=100.0,
                    upper quartile=100.0,
                    lower quartile=84.61538461538461,
                    upper whisker=100.0,
                    lower whisker=84.61538461538461
                },
                fill=cyan!40, draw=black,
            ] coordinates {};
    
            \addplot [only marks, mark=*, mark size=1.5pt] coordinates {(5,0.0) (5,0.0)};
    
            \addplot+ [
                boxplot prepared={
                    median=100.0,
                    upper quartile=100.0,
                    lower quartile=84.61538461538461,
                    upper whisker=100.0,
                    lower whisker=84.61538461538461
                },
                fill=cyan!40, draw=black,
            ] coordinates {};
    
            \addplot [only marks, mark=*, mark size=1.5pt] coordinates {(6,0.0) (6,0.0)};
    
            \addplot+ [
                boxplot prepared={
                    median=100.0,
                    upper quartile=100.0,
                    lower quartile=84.61538461538461,
                    upper whisker=100.0,
                    lower whisker=84.61538461538461
                },
                fill=cyan!40, draw=black,
            ] coordinates {};
    
            \addplot [only marks, mark=*, mark size=1.5pt] coordinates {(7,0.0) (7,0.0)};
    
            \addplot+ [
                boxplot prepared={
                    median=100.0,
                    upper quartile=100.0,
                    lower quartile=84.61538461538461,
                    upper whisker=100.0,
                    lower whisker=84.61538461538461
                },
                fill=cyan!40, draw=black,
            ] coordinates {};
    
            \addplot [only marks, mark=*, mark size=1.5pt] coordinates {(8,0.0) (8,0.0)};
    
        \nextgroupplot[
            cycle list={{black}},
            title={Critical},
            xmin=0,
            xmax=9,
            xlabel={},
            xtick={1, 2, 3, 4, 5, 6, 7, 8},
            xticklabels={CMP, CFF, DCI, DP, SIMP, SA, SD, SS},
            xticklabel style={rotate=45, anchor=east},
            yticklabels={}
        ]
    
            \addplot+ [
                boxplot prepared={
                    median=76.875,
                    upper quartile=98.80952380952381,
                    lower quartile=42.5,
                    upper whisker=100.0,
                    lower whisker=0.0
                },
                fill=cyan!40, draw=black,
            ] coordinates {};
    
            \addplot+ [
                boxplot prepared={
                    median=71.875,
                    upper quartile=100.0,
                    lower quartile=27.5,
                    upper whisker=100.0,
                    lower whisker=0.0
                },
                fill=cyan!40, draw=black,
            ] coordinates {};
    
            \addplot+ [
                boxplot prepared={
                    median=100.0,
                    upper quartile=100.0,
                    lower quartile=53.75,
                    upper whisker=100.0,
                    lower whisker=0.0
                },
                fill=cyan!40, draw=black,
            ] coordinates {};
    
            \addplot+ [
                boxplot prepared={
                    median=71.875,
                    upper quartile=98.80952380952381,
                    lower quartile=27.5,
                    upper whisker=100.0,
                    lower whisker=0.0
                },
                fill=cyan!40, draw=black,
            ] coordinates {};
    
            \addplot+ [
                boxplot prepared={
                    median=65.47619047619048,
                    upper quartile=98.4375,
                    lower quartile=27.5,
                    upper whisker=100.0,
                    lower whisker=0.0
                },
                fill=cyan!40, draw=black,
            ] coordinates {};
    
            \addplot+ [
                boxplot prepared={
                    median=94.49404761904762,
                    upper quartile=100.0,
                    lower quartile=65.0,
                    upper whisker=100.0,
                    lower whisker=60.0
                },
                fill=cyan!40, draw=black,
            ] coordinates {};
    
            \addplot [only marks, mark=*, mark size=1.5pt] coordinates {(6,0.0) (6,0.0)};
    
            \addplot+ [
                boxplot prepared={
                    median=76.875,
                    upper quartile=98.80952380952381,
                    lower quartile=42.5,
                    upper whisker=100.0,
                    lower whisker=0.0
                },
                fill=cyan!40, draw=black,
            ] coordinates {};
    
            \addplot+ [
                boxplot prepared={
                    median=71.875,
                    upper quartile=98.80952380952381,
                    lower quartile=27.5,
                    upper whisker=100.0,
                    lower whisker=0.0
                },
                fill=cyan!40, draw=black,
            ] coordinates {};
        \end{groupplot}
        \end{tikzpicture}
        \label{fig:vdl_single_plugin_in_the_wild_bearer}
    }
    \refstepcounter{figure}
    \caption{Distribution of VDL for Njsscan and Bearer on the in-the-wild dataset, using single-plugin obfuscations. Results are grouped by each tool's internal severity levels.}
    \label{fig:vdl_single_plugin_in_the_wild}
\end{figure} 
Figure~\ref{fig:vdl_single_plugin_in_the_wild} presents the VDL distributions when applying each obfuscation technique in isolation, with results grouped by the vulnerability severity reported by each tool.

\paragraph{Njsscan (Figure~\ref{fig:vdl_single_plugin_in_the_wild_njsscan})}
Njsscan's results on the in-the-wild dataset confirm its extreme fragility. The analysis is split by its two severity levels: 'Warning' and 'Error'. For vulnerabilities classified as 'Error', every single obfuscation technique achieves a median VDL of 100\%. The near-total absence of variance indicates that any single transformation is sufficient to reliably conceal these high-priority findings from the tool.

For 'Warning'-level vulnerabilities, the median VDL remains at or near 100\% for most techniques (e.g., \textit{SS}, \textit{DP}, \textit{SD}, \textit{SA}). We observe slightly more variance and a large number of outliers for techniques like CFF and DCI. This suggests that while some lower-priority detections may occasionally persist, the tool's overall effectiveness is still compromised.

\paragraph{Bearer (Figure~\ref{fig:vdl_single_plugin_in_the_wild_bearer})}
Bearer's performance is more nuanced and reveals a strong dependency on vulnerability severity. For 'High', 'Critical', and 'Low' severity vulnerabilities, the median VDL for all tested techniques is at or near 100\%. This indicates a widespread failure in detection. However, the presence of dense outlier clusters, particularly in the 'High' and 'Low' categories, points to an \textbf{'all-or-nothing' phenomenon}: while the median project's vulnerabilities are completely hidden, a substantial number of other projects are unaffected (0\% VDL).

The most significant finding concerns the \textbf{'Medium'} severity category. Here, the VDL distributions are much wider, and medians fall well below 100\%. For instance, \textit{CFF} and DCI show median VDLs of approximately 75\% and 30\%, respectively, with wide interquartile ranges. This demonstrates that Bearer's heuristics for 'Medium' severity issues are fundamentally more resilient and are not as easily defeated as those for other categories.

\begin{tcolorbox}[colback=gray!5,colframe=blue!40, boxrule=0.3mm]
\textbf{RQ2:} The impact of individual techniques is severe and highly dependent on vulnerability severity. For Njsscan, all techniques are devastating, achieving 100\% median VDL for 'Error' vulnerabilities. For Bearer, techniques like SA also achieve 100\% median VDL for 'High', 'Low', and 'Critical' issues. However, their impact is notably reduced against 'Medium' severity vulnerabilities, where Bearer retains partial detection capabilities. [See Figure~\ref{fig:vdl_single_plugin_in_the_wild}]
\end{tcolorbox}

Having analyzed the impact of individual techniques (RQ2), we now investigate the cumulative effect of stacking multiple obfuscations, as posed in RQ3. We assess whether increasing the number of applied transformations yields a correspondingly significant increase in VDL, or if the tools' failures plateau after a certain complexity. 

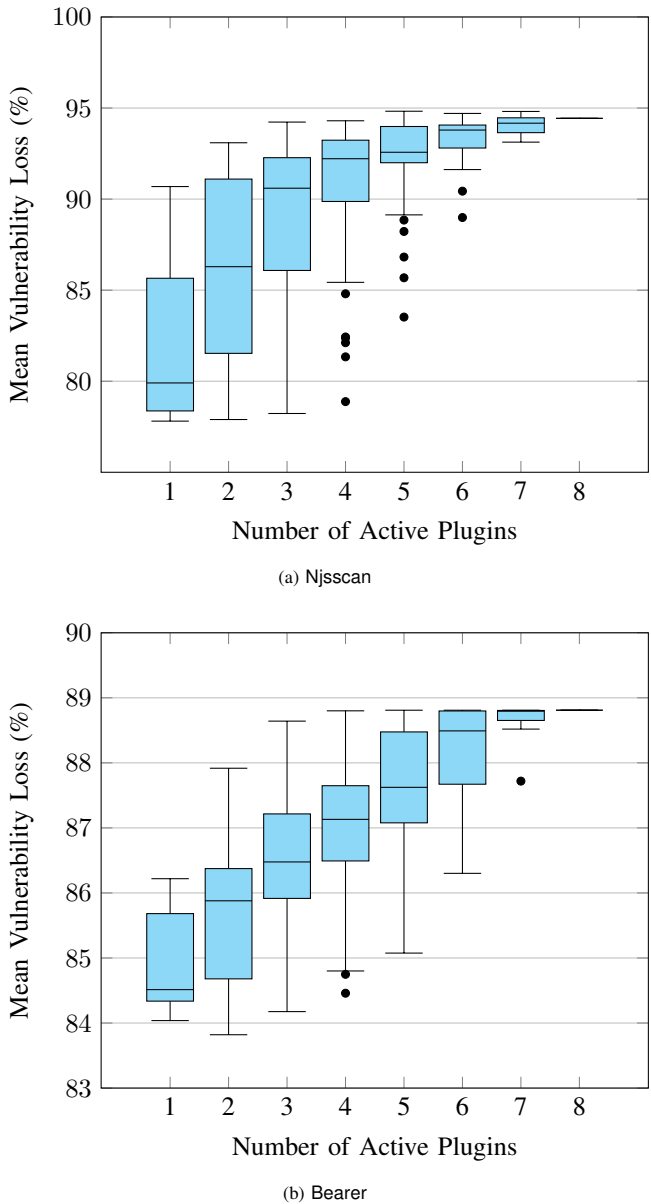
\begin{figure}
    \captionsetup[subfloat]{font=small, labelfont=scriptsize}
    \subfloat[\scriptsize Njsscan]{
    \begin{tikzpicture}
    \begin{axis}
    [
        cycle list={{black}},
        width=\columnwidth,
        ymin=75,
        ymax=100,
        ytick={80, 85, 90, 95, 100},
        boxplot/draw direction=y,
        ymajorgrids,
        xlabel={Number of Active Plugins},
        ylabel={Mean Vulnerability Loss (\%)},
        ylabel style={yshift=-5pt},
        xtick={1, 2, 3, 4, 5, 6, 7, 8},
        xticklabels={1, 2, 3, 4, 5, 6, 7, 8}
    ]

        \addplot+ [
            boxplot prepared={
                median=79.90490524219211,
                upper quartile=85.6533656398404,
                lower quartile=78.36843643352253,
                upper whisker=90.68796122250069,
                lower whisker=77.8078933462551
            },
            fill=cyan!40, draw=black,
        ] coordinates {};

        \addplot+ [
            boxplot prepared={
                median=86.28821419234109,
                upper quartile=91.09911922720312,
                lower quartile=81.52951150711509,
                upper whisker=93.09311991562831,
                lower whisker=77.89592151526918
            },
            fill=cyan!40, draw=black,
        ] coordinates {};

        \addplot+ [
            boxplot prepared={
                median=90.59763623052588,
                upper quartile=92.27331220929511,
                lower quartile=86.08101161868827,
                upper whisker=94.22885593938226,
                lower whisker=78.22721483071741
            },
            fill=cyan!40, draw=black,
        ] coordinates {};

        \addplot+ [
            boxplot prepared={
                median=92.21666383747367,
                upper quartile=93.23434888476754,
                lower quartile=89.86853244762872,
                upper whisker=94.30010861561813,
                lower whisker=85.42859876426506
            },
            fill=cyan!40, draw=black,
        ] coordinates {};

        \addplot [only marks, mark=*, mark size=1.5pt] coordinates {(4,82.11439757631382) (4,82.38337253719018) (4,84.80080816218035) (4,81.33805563692782) (4,82.42886565853074) (4,78.8796663198577)};

        \addplot+ [
            boxplot prepared={
                median=92.5716242821506,
                upper quartile=93.98640463557577,
                lower quartile=91.99648580702612,
                upper whisker=94.82443459840076,
                lower whisker=89.13354883420674
            },
            fill=cyan!40, draw=black,
        ] coordinates {};

        \addplot [only marks, mark=*, mark size=1.5pt] coordinates {(5,88.22342720979938) (5,83.52152300118809) (5,88.84828548644337) (5,86.81784886018295) (5,85.67899910425106)};

        \addplot+ [
            boxplot prepared={
                median=93.79083719380273,
                upper quartile=94.06554134273858,
                lower quartile=92.80408995707364,
                upper whisker=94.70413654910223,
                lower whisker=91.62059074544875
            },
            fill=cyan!40, draw=black,
        ] coordinates {};

        \addplot [only marks, mark=*, mark size=1.5pt] coordinates {(6,90.43539461601662) (6,88.98830816218035)};

        \addplot+ [
            boxplot prepared={
                median=94.16790505099254,
                upper quartile=94.46314648517176,
                lower quartile=93.64788251205664,
                upper whisker=94.812529836496,
                lower whisker=93.12751910185193
            },
            fill=cyan!40, draw=black,
        ] coordinates {};

        \addplot+ [
            boxplot prepared={
                median=94.43345715479174,
                upper quartile=94.43345715479174,
                lower quartile=94.43345715479174,
                upper whisker=94.43345715479174,
                lower whisker=94.43345715479174
            },
            fill=cyan!40, draw=black,
        ] coordinates {};

    \end{axis}
    \end{tikzpicture}
    \label{fig:vdl_by_plugin_number_in_the_wild_njsscan}
    }

    \subfloat[\scriptsize Bearer]{
    \begin{tikzpicture}
    \begin{axis}
    [
        cycle list={{black}},
        width=\columnwidth,
        boxplot/draw direction=y,
        ymajorgrids,
        ymin=83,
        ymax=90,
        ytick={80, 81, 82, 83, 84, 85, 86, 87, 88, 89, 90},
        xlabel={Number of Active Plugins},
        ylabel={Mean Vulnerability Loss (\%)},
        ylabel style={yshift=-5pt},
        xtick={1, 2, 3, 4, 5, 6, 7, 8},
        xticklabels={1, 2, 3, 4, 5, 6, 7, 8},
    ]

        \addplot+ [
            boxplot prepared={
                median=84.51349340712414,
                upper quartile=85.68232333621313,
                lower quartile=84.33657801963749,
                upper whisker=86.21902197468276,
                lower whisker=84.03756871953208
            },
            fill=cyan!40, draw=black
        ] coordinates {}; 

        \addplot+ [
            boxplot prepared={
                median=85.8792882061764,
                upper quartile=86.37439215572782,
                lower quartile=84.67915166721018,
                upper whisker=87.91743487512126,
                lower whisker=83.82002010520826
            },
            fill=cyan!40, draw=black
        ] coordinates {}; 

        \addplot+ [
            boxplot prepared={
                median=86.47668460486551,
                upper quartile=87.21530491322976,
                lower quartile=85.91624713971044,
                upper whisker=88.64168314766495,
                lower whisker=84.17561952927518
            },
            fill=cyan!40, draw=black
        ] coordinates {}; 

        \addplot+ [
            boxplot prepared={
                median=87.13140037177384,
                upper quartile=87.64889053620124,
                lower quartile=86.4921117147541,
                upper whisker=88.800032211812,
                lower whisker=84.8004580615742
            },
            fill=cyan!40, draw=black
        ] coordinates {}; 

        \addplot [only marks, mark=*, mark size=1.5pt] coordinates {(4,84.74884761339959) (4,84.45829301452653)};

        \addplot+ [
            boxplot prepared={
                median=87.62386512066519,
                upper quartile=88.47586528225867,
                lower quartile=87.07779124319886,
                upper whisker=88.80946261543328,
                lower whisker=85.07533732559716
            },
            fill=cyan!40, draw=black
        ] coordinates {}; 

        \addplot+ [
            boxplot prepared={
                median=88.49169157599746,
                upper quartile=88.79802425222235,
                lower quartile=87.67158634793614,
                upper whisker=88.80946261543328,
                lower whisker=86.30138029963045
            },
            fill=cyan!40, draw=black
        ] coordinates {}; 

        \addplot+ [
            boxplot prepared={
                median=88.79418867086355,
                upper quartile=88.80599965285751,
                lower quartile=88.65201914631582,
                upper whisker=88.80946261543328,
                lower whisker=88.51918326932937
            },
            fill=cyan!40, draw=black
        ] coordinates {}; 

        \addplot [only marks, mark=*, mark size=1.5pt] coordinates {(7,87.71928485132804)};

        \addplot+ [
            boxplot prepared={
                median=88.80946261543328,
                upper quartile=88.80946261543328,
                lower quartile=88.80946261543328,
                upper whisker=88.80946261543328,
                lower whisker=88.80946261543328
            },
            fill=cyan!40, draw=black
        ] coordinates {}; 

    \end{axis}
    \end{tikzpicture}
    
    \label{fig:vdl_by_plugin_number_in_the_wild_bearer}
    }
    \refstepcounter{figure}
    \caption{VDL distribution on the in-the-wild dataset, grouped by the number of simultaneously active plugins.}
    \label{fig:vdl_by_plugin_number_in_the_wild}
\end{figure} 
Figure~\ref{fig:vdl_by_plugin_number_in_the_wild} plots the VDL distribution as a function of the number of simultaneously active obfuscation plugins used to generate each variant.

\paragraph{Njsscan (Figure~\ref{fig:vdl_by_plugin_number_in_the_wild_njsscan})}
Njsscan's results show a clear correlation between VDL and the number of active plugins, but with significant diminishing returns. With a single plugin, the median VDL is already high at approximately 80\%, though the interquartile range is wide. The median VDL rises sharply as the number of techniques increases to 2, 3, and 4. 

Critically, we observe a distinct plateau starting at 5 active plugins. From 5 to 8 stacked techniques, the median VDL remains stable at approximately 93\%-94\%, and the interquartile range becomes extremely compressed. This finding suggests that while Njsscan is somewhat resilient to 1 or 2 techniques, its analysis is consistently defeated by any combination of 5 or more, and adding further complexity yields no additional benefit for an attacker.

\paragraph{Bearer (Figure~\ref{fig:vdl_by_plugin_number_in_the_wild_bearer})}
Bearer's results show a similar plateau effect, but the implications are different. It is crucial to note the highly compressed Y-axis (83\%-90\%). The median VDL for a single plugin is already $\approx$85\%, which is significantly higher than Njsscan's starting point.

As more techniques are stacked, the median VDL rises slowly, plateauing at $\approx$88\% from 5 plugins onward. The key insight here is that the total VDL gain from stacking (from 1 to 8 plugins) is only $\approx$3 percentage points. This indicates that Bearer's detection logic is already substantially broken by the average single technique, and stacking provides only a marginal, albeit consistent, increase in evasion. The same 5-plugin plateau is observed, but the "all-or-nothing" behavior seen in Figure~\ref{fig:vdl_single_plugin_in_the_wild_bearer} is averaged out, resulting in a high, stable VDL.

\begin{tcolorbox}[colback=gray!5,colframe=blue!40, boxrule=0.3mm]
\textbf{RQ3:} There is a clear cumulative effect, but it plateaus quickly. For both Njsscan and Bearer, VDL increases with the number of stacked techniques, but we observe a strong point of diminishing returns. The median VDL and consistency (i.e., a shrinking interquartile range) stabilize after 5 active plugins. This indicates that stacking more than 5 techniques provides no significant additional evasion advantage. [See Figure~\ref{fig:vdl_by_plugin_number_in_the_wild}]
\end{tcolorbox}

 \section{Discussion \& Limitations}
\label{sec:discussion}
In this section, we contextualize the results of our analysis presented in Section~\ref{sec:experiments}, analyze why commonplace JavaScript obfuscations undermine current SAST techniques, articulate the practical implications for CI/CD workflows, and delineate the study’s limitations and avenues for future research.

\paragraph{Obfuscation induces substantial detection loss across tools and datasets} The empirical results show that applying a single obfuscation technique is typically sufficient to suppress the vast majority of vulnerability findings, and that this effect holds on both the \textbf{Ground-Truth} and \textbf{In-the-Wild} datasets. Worringly, this effect holds also for severe issues (Njsscan "Error"; Bearer "High/Critical"). These observations are consistent with the per-project VDL collapses on the benchmark dataset (\eg Njsscan reaching 100\% VDL on multiple vulnerable-by-design apps) and with the VDL distributions for single obfuscation plugin, where medians consistently approach 100\% for most techniques and severities. 

\paragraph{Stacking multiple obfuscation techniques produces diminishing returns} Although VDL increases as more obfuscations are combined, both tools reach a clear plateau around five active plugins, after which additional complexity yields little to no extra evasion advantage; for Bearer, the overall VDL gain from one to eight plugins is less than 5\% because a single technique already breaks most detections. 

\paragraph{Rule-based and pattern-based analyses explain the fragility under semantics-preserving transformations} The studied tools primarily rely on AST-level patterns, shallow data-flow or control-flow reasoning, and rule matching. Semantics-preserving obfuscations alter surface structure (\eg renamings, layout simplifications), reshape control skeletons (\eg flattening, opaque branches), and rewrite data (\eg string arrays/splitting), thereby decoupling vulnerable source–sink relations from the syntactic cues the rules expect. This mismatch accounts for the near-universal failures observed under single-plugin transformations.

\paragraph{Near-term and long-term countermeasures can improve robustness} In the near term, pipelines should (i) detect and flag obfuscation to set expectations, (ii) apply safe normalization/deobfuscation where feasible (e.g., demangling, minification-aware parsing, structural simplification), and (iii) complement SAST with DAST or fuzzing for critical entry points. In the long term, research should focus on finding novel SAST methodologies able to withstand obfuscation, going beyond classical pattern-matching or other syntactic approaches towards more in-depth analysis with a semantical understanding of the analyzed code. These could be complemented leveraging recent GenAI-based approaches to code vulnerability detection~\cite{sheng2025llms} in order to produce more robust SAST tools.

\paragraph{Threats to validity} First, the tooling scope threatens external validity: we focus on open-source SAST and a representative open-source obfuscator; proprietary engines or obfuscators may behave differently. Second, our lens is static: while DAST is less sensitive to purely semantics-preserving transformations, its effectiveness remains constrained by coverage and runtime anti-instrumentation, so it cannot serve as a complete substitute. Third, our metric is baseline-relative: VDL quantifies loss with respect to each tool’s clean-code detections, which means absolute nondiscovery may be higher for tools with weak baselines. Fourth, our datasets (one with ground truth and one in the wild) improve internal and external validity but still reflect particular language features, build systems, and dependency mixes that may bias outcomes. Fifth, our stacking analysis studies the number of plugins rather than all permutations, so order effects may exist. Sixth, operational issues (\eg empty or failed reports) occurred on a non-negligible subset of repositories; we report these qualitatively and avoid drawing quantitative conclusions from failed runs. Despite these limitations, the magnitude and consistency of VDL across settings support the central conclusion. 

\paragraph{Takeaway} The principal takeaway is that current JavaScript SAST is not robust to commonplace obfuscations. Unless pipelines incorporate obfuscation awareness and complementary testing, a clean SAST report on obfuscated code should not be interpreted as evidence of security.

 \section{Conclusions}
\label{sec:conclusions}
This paper provided the first systematic empirical assessment of how common JavaScript obfuscations undermine the effectiveness of state-of-practice SAST in modern DevSecOps workflows. We executed a two-phase study on two different web app datasets, a Ground-Truth dataset composed of vulnerable-by-design web applications and a In-the-Wild dataset composed of Node.js and Javascript-based applications gathered from GitHub. The objective of our study was to quantify vulnerability detection degradation for representative SAST tools under single and stacked obfuscations. 

Our results show that even a single semantics-preserving transformation typically suppresses the vast majority of baseline findings, including high-severity issues, across both datasets. Stacking additional techniques yields diminishing returns: VDL plateaus once approximately five plugins are active, indicating that today’s engines fail primarily because any syntactic distortion perturbs their rule- and pattern-based reasoning, rather than because of transformation complexity per se. Tool behavior differs in degree, not in kind: Njsscan exhibits systemic failure (with multiple benchmarks collapsing to 100\% VDL), while Bearer’s resilience is inconsistent and severity-dependent, retaining partial signal mainly for “Medium”-level findings. 

These findings have immediate implications for practice. First-party code should be scanned before any build-time obfuscation, and obfuscated third-party dependencies should be treated as out of scope for dependable static assurance unless pipelines compensate. In particular, CI/CD should (i) detect and flag obfuscation to set expectations, (ii) apply safe normalization (e.g., deobfuscation, minification-aware parsing) when feasible, and (iii) complement SAST with DAST/fuzzing for critical surfaces, recognizing the coverage limits of dynamic techniques. 

Our study also motivates concrete research directions. Robust SAST for web code will require obfuscation-invariant intermediate representations and normalization passes; stronger taint/traversal across flattened control flow; selective semantic lifting (symbolic or concolic assistance) to recover flows hidden by opaque constructs; and learning-assisted detectors trained adversarially on transformed code. Pursuing these directions promises to materially reduce VDL under the obfuscations we measured. 

In sum, current JavaScript SAST is not robust to commonplace obfuscations. Without obfuscation awareness and orthogonal testing, “clean” static reports over obfuscated code offer at best a false sense of security. We hope our metric, datasets, and empirical evidence provide a foundation for building the next generation of obfuscation-resilient analysis for the web ecosystem.
 
\bibliography{biblio}
\bibliographystyle{IEEEtran}

\newpage

\section{Biography Section}

\vspace{11pt}

\vspace{-40pt}
\begin{IEEEbiography}[{\includegraphics[width=1in,height=1.25in,clip,keepaspectratio]{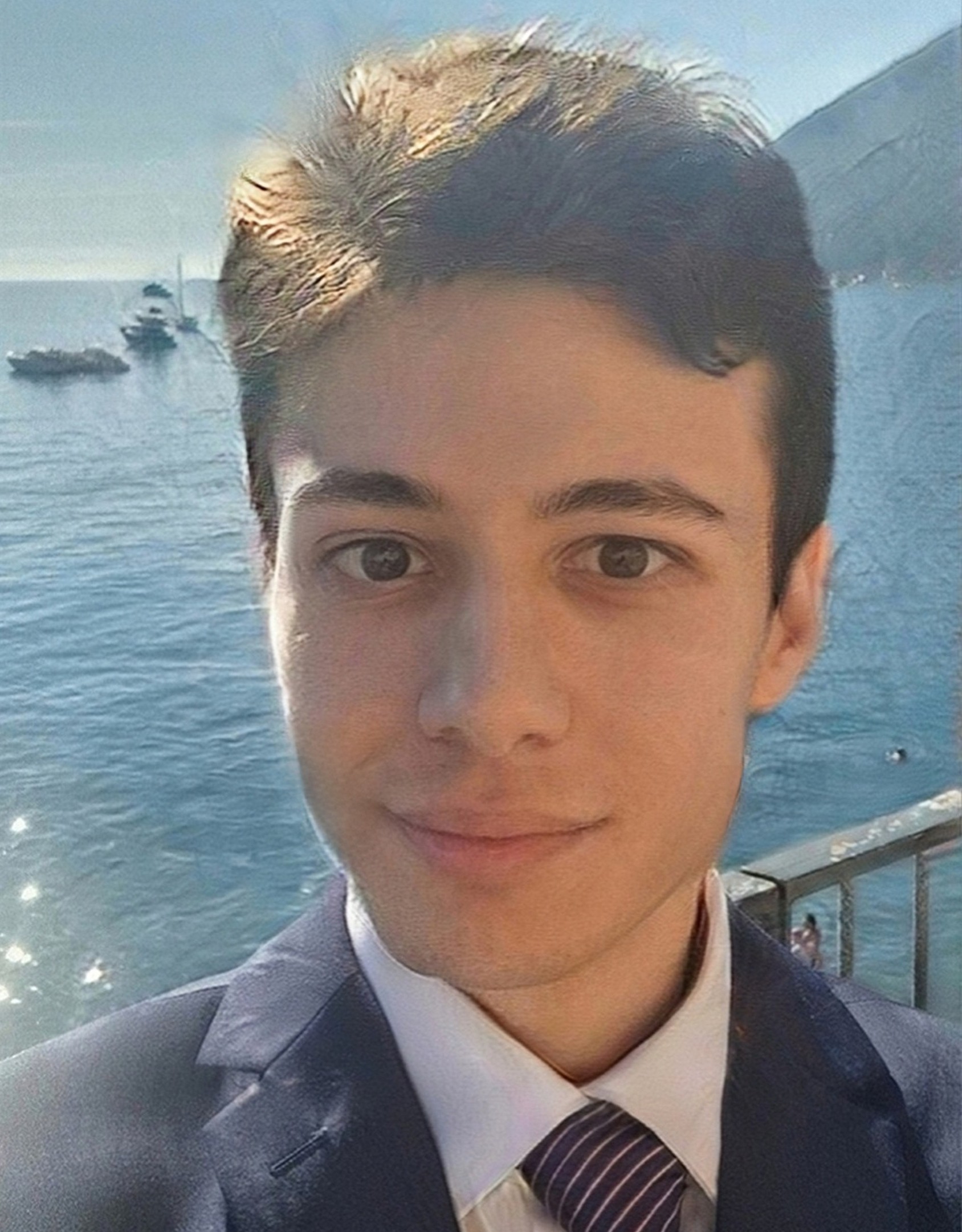}}]{Francesco Pagano} is a Postdoctoral Researcher at the University of Verona, Italy. He received the International PhD in Computer Engineering from the University of Genoa, Italy, in 2025. He was a visiting researcher for six months at the TAU Group, Università della Svizzera italiana (USI), Lugano. His research interests include software engineering and security, specifically vulnerability detection in mobile/IoT ecosystems and Android user privacy.
\end{IEEEbiography}

\vspace{-40pt}
\begin{IEEEbiography}[{\includegraphics[width=1in,height=1.25in,clip,keepaspectratio]{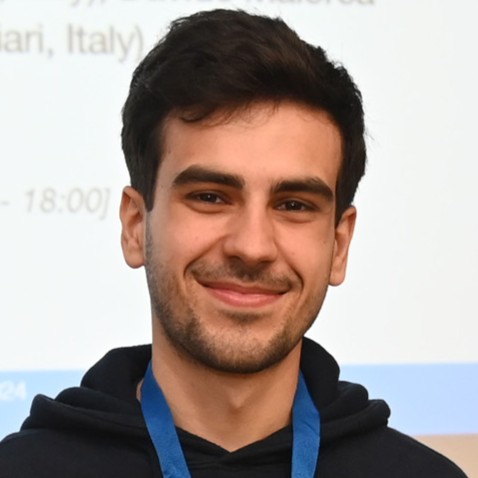}}]{Lorenzo Pisu} is a third-year PhD student in electronics and computer engineering at the University of Cagliari. He received a Master's degree in Computer Engineering, Cybersecurity, and Artificial Intelligence in 2022. His main research focuses on Web Security, specifically on recent vulnerabilities and technologies such as Server-Side Template Injection (SSTI) and HTTP/3.
\end{IEEEbiography}

\vspace{-40pt}
\begin{IEEEbiography}[{\includegraphics[width=1in,height=1.25in,clip,keepaspectratio]{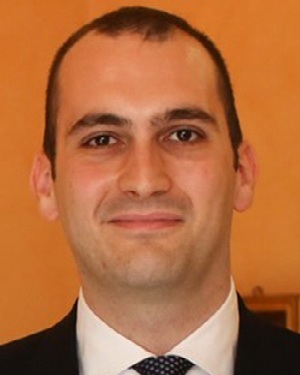}}]{Leonardo Regano} received an M.Sc. degree in 2015 and a Ph.D. in Computer Engineering in 2019 from Politecnico di Torino, where he worked as a research assistant for eight years. He is currently an assistant professor at the Department of Electrical and Electronic Engineering, University of Cagliari (Italy). His current research interests focus on software security, artificial intelligence and machine learning applications to cybersecurity, security policies analysis, and software protection techniques assessment. 
\end{IEEEbiography}

\vspace{-40pt}
\begin{IEEEbiography}[{\includegraphics[width=1in,height=1.25in,clip,keepaspectratio]{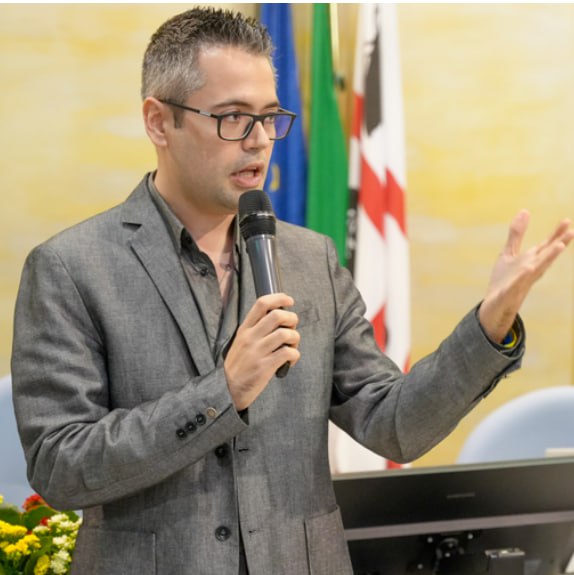}}]{Davide Maiorca}  received the M.S. degree (Electronic Engineering, 110/110 cum laude) in 2012 and the Ph.D. degree in Electronic Engineering and Computer Science in 2016, both from the University of Cagliari, Italy. He has been with the sAIfer Lab since 2012 and is currently an Associate Professor of Computer Engineering at the University of Cagliari.

His research interests include the analysis and detection of x86, Android, and IoT malware, the study of malicious documents and multimedia applications (such as PDF and Microsoft Office files), and Adversarial Machine Learning. He has authored more than 40 peer-reviewed publications and has served as a Program Committee member and reviewer for several international conferences and journals. Prof. Maiorca is a member of IEEE, the IEEE Computer Society, IEEE SMC, ACM, and CVPL (formerly GIRPR). 
\end{IEEEbiography}

\vspace{-40pt}
\begin{IEEEbiography}[{\includegraphics[width=1in,height=1.25in,clip,keepaspectratio]{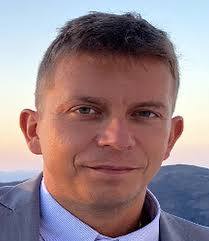}}] {Alessio Merlo} (Ph.D. 2010) is a full professor in Computer Engineering and the Director of the CASD - University School of Advanced Defense Studies in Rome. He has been a researcher on cybersecurity since 2010, mainly focusing on mobile security, data privacy, and cyberphysical systems security. 
\end{IEEEbiography}

\vspace{-40pt}
\begin{IEEEbiography}[{\includegraphics[width=1in,height=1.25in,clip,keepaspectratio]{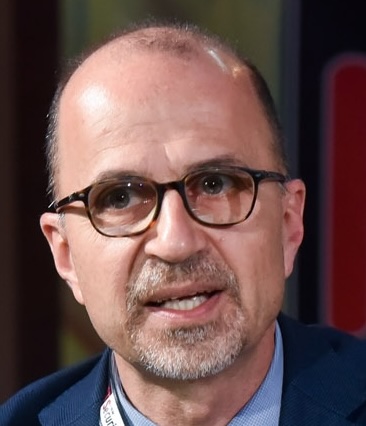}}] {Giorgio Giacinto}  is a Professor of Computer Engineering at the University of Cagliari, Italy. His research activities are carried out within the sAIfer Lab research group, where he leads the Cybersecurity research area. His main contributions lie in machine learning approaches to cybersecurity, specifically regarding threat analysis and detection, and are supported by funding from national and international projects. He has published nearly 200 papers in international conferences and journals and regularly serves as a member of the Editorial Board and Program Committee for several international journals and conferences.
He is a vice director of the Cybersecurity National Lab and a member of the managing committee of the Artificial Intelligence \& Intelligent Systems Lab, both within the CINI consortium in Italy. He is also a member of the expert board for the Italian delegation in the Program Committee of Horizon Europe, Cluster III on “Civil Security for Society.”
He is a Fellow of the IAPR (International Association for Pattern Recognition) and a Senior Member of the IEEE Computer Society and ACM.
\end{IEEEbiography}

\vfill

\end{document}